\documentclass[useAMS,usenatbib,a4paper]{mn2e}

\usepackage[utf8]{inputenc} 
\usepackage[T1]{fontenc}    
\usepackage{hyperref}       
\usepackage{url}            
\usepackage{booktabs}       
\usepackage{amsfonts}       
\usepackage{nicefrac}       
\usepackage{microtype}      
\usepackage{lipsum}
\usepackage{fancyhdr}       
\usepackage{graphicx}       

\usepackage{cuted}
\usepackage{subcaption}
\usepackage{dsfont}
\usepackage{multirow}
\usepackage{natbib,fancyhdr}
\usepackage{bm}
\usepackage{amsmath}
\usepackage{amssymb}
\usepackage{mathtools}
\usepackage{orcidlink}
\usepackage{tikz}
\usetikzlibrary{calc}
\usetikzlibrary{automata}
\usetikzlibrary{positioning}  
\usetikzlibrary{arrows}   
\tikzset{stretch/.initial=1}    
\newcommand\drawloop[4][]%
   {\draw[shorten <=0pt, shorten >=0pt,#1]
      ($(#2)!\pgfkeysvalueof{/tikz/stretch}!(#2.#3)$)
      let \p1=($(#2.center)!\pgfkeysvalueof{/tikz/stretch}!(#2.north)-(#2)$),
          \n1= {veclen(\x1,\y1)*sin(0.5*(#4-#3))/sin(0.5*(180-#4+#3))}
      in arc [start angle={#3-90}, end angle={#4+90}, radius=\n1]%
   }
\usepackage{tikz-cd}
\usetikzlibrary{cd}

\usepackage{aas_macros} 

\newcommand{\Esp}[0]{\ensuremath{\mathbb{E}}}
\newcommand{\DKL}[0]{\ensuremath{\mathbb{D}_{\mathsf{KL}}}}
\newcommand{\nn}{\nonumber}

\title[Galaxy Imaging with Generative Models]{Galaxy Imaging with Generative Models: Insights from a Two-Models Framework}

\author[J.E Campagne]{
 Jean-Eric Campagne\thanks{E-mail: jean-eric.campagne@ijclab.in2p3.fr}\orcidlink{0000-0002-1590-6927}\\
 Université Paris-Saclay, CNRS/IN2P3, IJCLab,  91405 Orsay, France
}

\pubyear{2025}

\begin{document}
\label{firstpage}
\pagerange{\pageref{firstpage}--\pageref{lastpage}}
\maketitle

\begin{abstract}
Generative models have recently revolutionized image generation tasks across diverse domains, including galaxy image synthesis. This study investigates the statistical learning and consistency of three generative models: \texttt{light-weight-gan} (a GAN-based model), \texttt{Glow} (a Normalizing Flow-based model), and a diffusion model based on a \texttt{U-Net} denoiser, all trained on non-overlapping subsets of the SDSS dataset of $64 \times 64$ grayscale images. While all models produce visually realistic images with well-preserved morphological variable distributions, we focus on their ability to learn and generalize the underlying data distribution.

The diffusion model shows a transition from memorization to generalization as the dataset size increases, confirming previous findings. Smaller datasets lead to overfitting, while larger datasets enable novel sample generation, supported by the denoising process's theoretical basis. For the flow-based model, we propose an "inversion test" leveraging its bijective nature. Similarly, the GAN-based model achieves comparable morphological consistency but lacks bijectivity. We then introduce a "discriminator test," which shows successful learning for larger datasets but poorer confidence with smaller ones. Across all models, dataset sizes below $O(10^5)$ pose challenges to learning.

Along our experiments, the "two-models" framework enables robust evaluations, highlighting both the potential and limitations of these models. These findings provide valuable insights into statistical learning in generative modeling, with applications certainly extending beyond galaxy image generation.
\end{abstract}


\section{Introduction}
\label{sec:Intro}
Image generation in Machine Learning (ML) is a challenging task, particularly in the context of non-ergodic processes. Recently, significant breakthroughs in image quality have been achieved thanks to large-scale statistical model architectures such as \texttt{DALL-E2} \citep{ramesh2022}, \texttt{Midjourney} \citep{Oppenlaender2022}, and \texttt{StableDiffusion} \citep{Rombach2022}, which leverage \textit{stochastic diffusion processes}. These models have largely replaced the previous generation of approaches based on \textit{variational autoencoders} (VAE) \citep{Kingma2014}\footnote{Note: Updated in 2022.}, \textit{adversarial networks} (GAN) \citep{goodfellow2014generative} as employed in works such as \citep[e.g.,][]{KarrasALL18,Brock2019}, and \textit{normalizing flows}, exemplified by \texttt{Glow} \citep{Kingma2018}. 

Generative models have rapidly found applications across diverse domains. For instance, in High Energy Physics, a comprehensive review can be found in \cite{PhysRevD.107.076017}. Regarding galaxy image generation, several architectures have been explored, often in conjunction with deblending and deconvolution tasks. For example, \cite{ravanbakhsh2016} employed conditional VAE and GAN models, while \citep{Schawinski2017,Fussell2019,Hemmati_2022} utilized GANs. \cite{Arcelin2020} adopted a VAE-based approach, whereas \cite{Lanusse2021} proposed a hybrid VAE-Normalizing Flow architecture. Recently, \cite{smith2021} introduced a denoising stochastic diffusion model. These generative models aim to surpass parameterized analytical light profile simulations, such as those provided by \texttt{GalSim} \citep{ROWE2015121}, by capturing complex structures and interactions between background signals and those of single or blended galaxies. 

Despite the remarkable image quality achieved by modern generative models, which are widely accessible to the general public, several critical questions remain. These pertain to both mathematical foundations and practical applications. A key issue involves understanding what generative models truly learn and the statistical properties of their outputs. For example, \cite{Hataya2023} raised concerns about whether generated images could corrupt future datasets. Their study, however, focused on large-scale statistical models typically trained on billion-scale datasets sourced from the internet, which are prone to contamination by widely shared user-generated content. In contrast, galaxy image datasets remain relatively modest in size (i.e., $O(10^5)$), sourced from optical surveys such as COSMOS (HST Advanced Camera for Surveys; \cite{mandelbaum_2019_3242143}) and the Sloan Digital Sky Survey \citep[SDSS;][]{sdss}. These datasets, used by a smaller community compared to social media, are less susceptible to such contamination.

To address the fidelity of generated galaxy images and their morphological properties relative to original datasets, \cite{HACKSTEIN2023100685} and \cite{janulewicz2024assessing} have investigated various metrics. Furthermore, \cite{kadkhodaie2024generalization} tackled more mathematically oriented questions. Notably, they demonstrated the transition from memorization to generalization in diffusion generative models as dataset size increases. They also proposed an interpretation of what these networks learn, characterizing it as a form of \textit{geometry-adaptive harmonic basis}, extending beyond the commonly used {steerable wavelet basis}. 

To achieve this, \cite{kadkhodaie2024generalization} employed \textit{denoiser} architectures, such as \texttt{U-Net} \citep{ronneberger2015u} and \texttt{BF-CNN} networks \citep{Mohan2020Robust}, trained on reduced datasets like \texttt{CelebA} \citep{Liu2015} and \texttt{LSUN} Bedroom \citep{Yu2015}, comprising $O(10^5)$ grayscale images resized to $80 \times 80$ pixels.  This dataset scale enables a systematic investigation of how various generative models, including normalizing flows, GANs, and denoiser-based diffusion models, perform in galaxy image generation when trained on the same dataset derived from the SDSS survey. 

In the following sections, we begin by briefly describing various types of generative models, including Variational Auto-Encoders, Generative Adversarial Networks, Normalizing Flows, and Score-based Diffusion Models (Section~\ref{sec-generative-models}). Next, we present the results of our numerical experiments (Section~\ref{sec-experiment}), which primarily involve pairs of models with identical architectures trained on non-overlapping datasets of the same size. This approach not only allow comparisons of morphological variable distributions, a common practice for validating the similarity between generated and real data samples using a single model, but also enables the implementation of consistency tests with two models. Finally, we provide our conclusions and discuss potential avenues for future research (Section~\ref{sec-conclusion}). We provide the material to replay our experiment at the following location \url{https://github.com/jecampagne/galaxy-gen-model-compagnon}.
\section{Generative Models}
\label{sec-generative-models}

Generative models assume that observations $\bm{x} \in \mathbb{R}^d$ (where $d$ represents the number of pixels times the number of channels in image modeling) are described by a probability distribution $p(\bm{x})$, and aim to approximate this distribution. It is further assumed that the dataset used for training consists of i.i.d. samples drawn from $p(\bm{x})$. In this section, we provide background knowledge on generative models rather than a detailed description of their architectures or implementations. We begin with the Variational Auto-Encoder (VAE), mainly for the sake of completeness and to introduce relevant terminology, although this architecture is not included in our current experiments (Section~\ref{sec-experiment}), which focus on GANs, normalizing flows, and score-based diffusion models. Details on some implementations used are described in Section~\ref{sec-Exp-Models}.

\subsection{Variational Auto-Encoder}

The \texttt{VAE} architecture \citep{Kingma2014} assumes the existence of an underlying unobservable stochastic variable $\bm{z} \in \mathbb{R}^{d_\ell}$ (commonly referred to as a \textit{latent variable}) whose distribution is selected \textit{a priori} from a parameterized family $\pi_{\bm{\theta}_1}(\bm{z})$ that is differentiable with respect to both $\bm{z}$ and $\bm{\theta}_1$. Typically, the latent space has a much lower dimensionality than the data space, implying information compression ($d_\ell \ll d$). The data distribution is then expressed as $p(\bm{x}) = \int p(\bm{x}|\bm{z}) \pi_{\bm{\theta}_1}(\bm{z}) \, d\bm{z}$. However, the likelihood $p(\bm{x}|\bm{z})$ is generally unknown, necessitating the introduction of a parameterized approximation $p_{\bm{\theta}_2}(\bm{x}|\bm{z})$ with similar differentiability properties. Using the notation $\bm{\theta} = (\bm{\theta}_1, \bm{\theta}_2)$, these parameters are optimized so that $p_{\bm{\theta}}(\bm{x})$ aligns with the empirical distribution of the $N$ training samples $\{\bm{x}^i\}_{i<N}$. The optimal parameters $\bm{\theta}_{ML}$ maximize $p_{\bm{\theta}}(\bm{x})$, but this requires computing the gradient of the integral 
\begin{equation}
\int p_{\bm{\theta}_2}(\bm{x}|\bm{z}) \pi_{\bm{\theta}_1}(\bm{z}) \, d\bm{z},
\end{equation}
which is often intractable both analytically and numerically \citep{Kingma2014}. 

To address this challenge, the Bayes rule is employed:
\begin{equation}
p_{\bm{\theta}}(\bm{x}) p(\bm{z}|\bm{x}) = p_{\bm{\theta}_2}(\bm{x}|\bm{z}) \pi_{\bm{\theta}_1}(\bm{z}),
\end{equation}
along with a parameterized approximation $p_{\bm{\phi}}(\bm{z}|\bm{x})$ for the unknown true \textit{a posteriori} distribution. Introducing the Kullback-Leibler (KL) divergence $\DKL(p\|q) = \mathbb{E}_{\bm{x} \sim p}[\log(p(\bm{x})/q(\bm{x}))]$, and taking the expected value according\footnote{We represent the sampling of variates $\bm{x}$ from a distribution $p(\bm{x})$ using the notation $\bm{x}\sim p(\bm{x})$.}  to $\bm{z} \sim p_{\bm{\phi}}(\bm{z}|\bm{x})$, we rewrite:
\begin{multline}
\log p_{\bm{\theta}}(\bm{x}) = \log p_{\bm{\theta}_2}(\bm{x}|\bm{z}) + \log \pi_{\bm{\theta}_1}(\bm{z}) - \log p(\bm{z}|\bm{x}) \\ 
+ \log p_{\bm{\phi}}(\bm{z}|\bm{x}) - \log p_{\bm{\phi}}(\bm{z}|\bm{x}),
\end{multline}
which leads to:
\begin{multline}
\log p_{\bm{\theta}}(\bm{x}) = \mathbb{E}_{\bm{z} \sim p_{\bm{\phi}}}[\log p_{\bm{\theta}_2}(\bm{x}|\bm{z})] 
- \DKL(p_{\bm{\phi}}(\bm{z}|\bm{x}) \| \pi_{\bm{\theta}_1}(\bm{z})) \\ 
+ \DKL(p_{\bm{\phi}}(\bm{z}|\bm{x}) \| p(\bm{z}|\bm{x})) \geq \mathcal{L}(\bm{x}; \{\bm{\theta}, \bm{\phi}\}),
\label{eq-VAE-ELBO}
\end{multline}
where we highlight the \textit{evidence lower bound} (ELBO):
\begin{equation}
\mathcal{L}(\bm{x}; \{\bm{\theta}, \bm{\phi}\}) = \mathbb{E}_{\bm{z} \sim p_{\bm{\phi}}}[\log p_{\bm{\theta}_2}(\bm{x}|\bm{z})] 
- \DKL(p_{\bm{\phi}}(\bm{z}|\bm{x}) \| \pi_{\bm{\theta}_1}(\bm{z})).
\end{equation}

Maximizing the ELBO with respect to $\{\bm{\theta}, \bm{\phi}\}$ is feasible because it eliminates dependence on the unknown distributions $p(\bm{x})$ and $p(\bm{z}|\bm{x})$. To mitigate the large variance of the gradient with respect to $\bm{\phi}$, \cite{Kingma2014} introduced an auxiliary variable $\varepsilon$, replacing $\bm{z} \sim p_{\bm{\phi}}(\bm{z}|\bm{x})$ with $\bm{z} = g_{\bm{\phi}}(\bm{\varepsilon}, \bm{x})$, where $\bm{\varepsilon} \sim p(\bm{\varepsilon})$ and $g_{\bm{\phi}}$ is a differentiable function with respect to $\bm{\phi}$.

In this framework, $p_{\bm{\theta}_2}(\bm{x}|\bm{z})$ is typically referred to as the \textit{decoder}, while $p_{\bm{\phi}}(\bm{z}|\bm{x})$ is the \textit{encoder}. Different VAE models vary in their choice of distribution families. A common choice is Gaussian multivariate distributions, with a centered isotropic prior of unit variance $\mathcal{N}(\bm{0}, \bm{1})$ for $\bm{z}$ (where $\bm{\theta}_1$ is omitted in this scenario). Once optimization is complete, the \textit{decoder} can generate $\bm{x}$ by sampling $\bm{z}$. However, simple VAEs often generate blurry images and may suffer from \textit{posterior collapse}, motivating ongoing research \citep[e.g.,][]{engel2018latent, Takida2022} and modified VAE architectures, such as those proposed by \cite{Lanusse2021}.

\subsection{Generative Adversarial Network}
\label{sec-GAN}
The vanilla GAN \citep{goodfellow2014generative} shares with VAE the purpose of optimizing a \textit{generator} network ($\bm{G}$) to obtain $\bm{x}$ samples from a latent variable $\bm{z}$ such that $\bm{x} = \bm{G}(\bm{z})$, with $\bm{z}\in\mathbb{R}^{d_\ell}$ (in general $d_\ell\ll d$). The \textit{prior} $\pi(\bm{z})$ is typically a Gaussian distribution $\mathcal{N}(\bm{0},\bm{1})$. A discussion on alternative choices of priors is presented by \cite{Brock2019}. Note that one can also use a prior fed by $\bm{z}\sim \mathcal{N}(\bm{0},\bm{1})$, which is learned during the generator optimization.

To optimize the generator, a second network ($\bm{D}$), acting as a \textit{discriminator}, aims to determine whether a sample $\bm{x}$ is from the model distribution or the dataset distribution. To achieve this, one solves the \texttt{min-max} problem ({Nash equilibrium}):
\begin{equation}
\min_{\bm{G}} \max_{\bm{D}}\left\{ \Esp_{\bm{x}\sim p_{data}(\bm{x})}[\log \bm{D}(\bm{x})] + \Esp_{\bm{z}\sim \pi(\bm{z})}[\log(1-\bm{D}(\bm{G}(\bm{z})))] \right\},
\end{equation}
where $p_{data}$ is the data-generating distribution. Both $\bm{G}$ and $\bm{D}$ are parameterized networks. 
{It can be shown that the \texttt{min-max} problem can be reformulated using the Jensen-Shannon metric to find the best generator $G$}
\begin{equation}
\min_G \left\{ \DKL{\left( p_{data} \left\| \frac{p_{data} + p_{\bm{G}}}{2}\right. \right)} + \DKL{\left(p_{\bm{G}} \left\| \frac{p_{data} + p_{\bm{G}}}{2} \right. \right)} \right\},
\end{equation}
yielding the optimal solution $p_{\bm{G}} = p_{data}$, which perfectly matches the dataset distribution.

However, training the GAN model via the objective function above is generally unstable, leading to a pathology referred to in the literature as \textit{mode-collapse}. This occurs because the \textit{discriminator} can overfit the dataset too quickly, causing vanishing gradients \citep{Gulrajani2017}. This training instability has been extensively studied \citep[see, e.g., reviews by][]{Saxena2021,Jozdani2022}. Architectures like \texttt{BigGAN} \citep{Brock2019} and \texttt{StyleGAN2} \citep{Karras2018ASG,Karras2020} address the gradient flow issue using various techniques, representing "state-of-the-art" improvements, albeit at the expense of computational resources.

\cite{liu2021towards} addressed training stability and reduced computational resource requirements, especially for small datasets, by developing a \texttt{light-weight-GAN} structure. One key element is the hinge loss introduced by \cite{Lim2017}, which minimizes the following losses to alternately train the \textit{discriminator}:
\begin{multline}
\min_{\bm{D}} \left\{\Esp_{\bm{x}\sim p_{data}}[\max{(0,1-\bm{D}(\bm{x}))}] 
\right. \label{eq-discri-hinge-loss}\\ 
 \left. + \Esp_{\bm{z}\sim \pi(\bm{z})}[\max{(0,1+\bm{D}(\bm{G}(\bm{z}))}]
\right\},
\end{multline}
and the \textit{generator}:
\begin{equation}
\min_{\bm{G}} \left\{-\Esp_{\bm{z}\sim \pi(\bm{z})}[\bm{D}(\bm{G}(\bm{z}))]
\right \}.
\end{equation}
Here, $\bm{D}$ is a linear discriminator, similar to those used in Support Vector Machines \citep{Vapnik1997}. The second key ingredient is a strong regularization applied to the \textit{discriminator} loss. The \textit{discriminator} $\bm{D}$ is treated as an \textit{encoder} (e.g., as in a VAE) and is trained alongside \textit{decoders} to ensure that $\bm{D}$ extracts image features at different scales (e.g., $8\times 8$, $16\times 16$) that allow the \textit{decoders} to reconstruct the images accurately. Finally, the \textit{generator} employs a new skip-layer excitation module (SLE), which enables a more robust gradient flow between feature maps extracted at different scales.

GAN architectures have been used for galaxy generation in conjunction with deblending and deconvolution tasks. For instance, \cite{Schawinski2017} and \cite{Hemmati_2022} employed vanilla-GANs, while \cite{Coccomini2021} utilized the \texttt{light-weight-GAN} that we use in our experiment too.
\subsection{Normalizing Flows}
\label{sec-NF}
Looking at Equation \ref{eq-VAE-ELBO}, one notices that the optimal solution is reached when $\DKL(p_{\bm{\phi}}(\bm{z}|\bm{x})\| p(\bm{z}|\bm{x})) = 0$, which is true if and only if $p_{\bm{\phi}}(\bm{z}|\bm{x}) = p(\bm{z}|\bm{x})$. However, this situation is generally unattainable because typical $p_{\bm{\phi}}$ parameterizations involve, for instance, independent Gaussian distributions. This limitation in parameterization is one of the drawbacks of VAEs. { These considerations motivate the use of normalizing flows (NF) rethinking the problem of probability transport from $\pi(\bm{z})$ to $p_{data}(\bm{x})$. Introducing more flexibility in the family of distributions, we can hope that the true posterior is a member of such a family \citep{Tabak2010, Tabak2013a, Rezende2015}.}

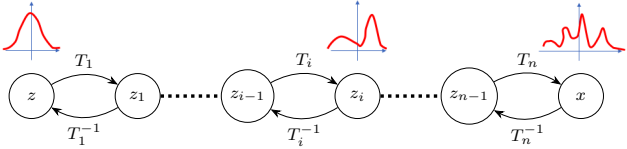
\begin{figure}
\begin{center}

\resizebox{\linewidth}{!}{%
\begin{tikzpicture}[
node distance = 1cm, 
distrib/.style={rectangle},
]

\node[state] (s0) {$z$};
\node[distrib, above=0.1cm of s0] (img0) {\includegraphics[height=1cm]{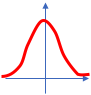}}; 
\node[state, right = of s0] (s1) {$z_1$};
\node[state, right = of s1] (sim1) {$z_{i-1}$};
\node[state, right = of sim1] (si) {$z_{i}$};
\node[distrib, above=0.1cm of si] (imgi) {\includegraphics[height=1cm]{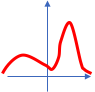}}; 
\node[state, right = of si] (snm1) {$z_{n-1}$};
\node[state, right = of snm1] (sn) {$x$};
\node[distrib, above=0.1cm of sn] (imgx) {\includegraphics[height=1cm]{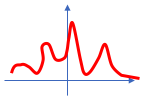}}; 

\path[-Stealth]
	(s0) edge[bend left] node [above] {$T_1$} (s1)
	(s1) edge[bend left] node [below] {$T_1^{-1}$} (s0)
	(s1) edge[dotted, ultra thick,-] (sim1)
	(sim1) edge[bend left] node [above] {$T_i$}  (si) 
	(si) edge[bend left] node [below] {$T_i^{-1}$}  (sim1) 
	(si) edge[dotted, ultra thick,-](snm1)
	(snm1) edge[bend left] node [above] {$T_n$}  (sn)
	(sn) edge[bend left] node [below] {$T_n^{-1}$}  (snm1); 

\end{tikzpicture}}
\caption{Schematic Normalizing Flow process. On the top the \textit{forward} or \textit{generative} direction from a simple $\bm{z}$ distribution to a more complex one for $\bm{x}$. On the bottom the \textit{backward}  or \textit{training} direction from complex to simple distributions.}
\label{fig-normflow}
\end{center}
\end{figure}

{
A flow $T$ is generally constructed as a composition of several individual diffeomorphism (i.e., a bijector)  $\{T_i\}_{i<n}$ such that:
\begin{align}
T = T_1 \circ T_2 \circ \dots \circ T_n & \quad \Leftrightarrow \quad 
T^{-1} = T_n^{-1} \circ T_{n-1}^{-1} \circ \dots \circ T_1^{-1}.
\end{align}
Unlike VAEs and GANs, the dimensionalities of the latent and data spaces are the same by definition in flow-based models.
Figure \ref{fig-normflow} illustrates a schematic view of the \textit{forward} or \textit{generative} direction and the \textit{backward/reverse} or \textit{training} direction. If we denote $\bm{z}_i = T_i(\bm{z}_{i-1})$ for all $i < n$ (using $\bm{z}_0 = \bm{z}$ and $\bm{z}_n = \bm{x}$), then:
\begin{equation}
\log |\det J_T| = \sum_{i=0}^{n-1} \log |\det J_{T_i}(\bm{z}_{i-1})|.
\label{eq-flow-jacob}
\end{equation}
}

Theorems exist that demonstrate, under moderate assumptions, flow-based models are capable of representing any density distribution \citep{Bogachev2005, huang2019solving}. Consequently, if the flows are well-designed, one can sample $\bm{x}$ with potentially complex (multi-modal) density distributions from a simple spherical centered multivariate Gaussian distribution.

Let the flow be parametrized by a vector $\bm{\theta}$, leading to a parametrized flow model $p_{\bm{\theta}}(\bm{x})$. To optimize the flow, we consider the $\DKL$ divergence between $p_{\bm{\theta}}(\bm{x})$ and the true $p(\bm{x})$, resulting in\footnote{For simplicity, we assume $\pi(\bm{z}) = \mathcal{N}(\bm{0}, \bm{1})$ without introducing additional unknown parameters.}:
\begin{align}
\mathcal{L}(\bm{\theta}) &= \DKL(p(\bm{x})\|p_{\bm{\theta}}(\bm{x})) 
= - \Esp_{\bm{x}\sim p(\bm{x})}[-\log p_{\bm{\theta}}(\bm{x})] + \mathrm{const.} \nn \\
&= - \Esp_{\bm{x}\sim p(\bm{x})}[\log \pi(T_{\bm{\theta}}^{-1}(\bm{x})) + \log |\det J_{T_{\bm{\theta}}^{-1}}(\bm{x})|] + \mathrm{const.}
\label{eq-NF-loss-1}
\end{align}
{ Using this method, we directly address the convergence of $p_\theta$ towards $p$ (i.e., $p_{\text{data}}$) using the Kullback-Leibler metric, while avoiding the GAN issue related to the \texttt{min-max} Nash equilibrium.}

{ In practice, computing the expectation $\Esp_{\bm{x} \sim p(\bm{x})}$ 
is performed by a Monte Carlo approach using the data samples $\{\bm{x}^{(i)}\}_{i<N}$.} 
Dropping the constant term, the loss function becomes:
\begin{equation}
\mathcal{L}(\bm{\theta}) = -\frac{1}{N} \sum_{i=0}^{N-1} \left\{ \log \pi(T_{\bm{\theta}}^{-1}(\bm{x}^{(i)})) + \log |\det J_{T_{\bm{\theta}}^{-1}}(\bm{x}^{(i)})| \right\}.
\label{eq-NF-loss-2}
\end{equation}

The computation of the determinant of the Jacobian and its gradient can be challenging. However, it becomes more straightforward if the Jacobian matrices $\{J_{T_i}^{-1}\}_{i<n}$ (Eq.~\ref{eq-flow-jacob}) are triangular. To minimize $\mathcal{L}(\bm{\theta})$ and obtain the best $\bm{\theta}$, we require $T_\theta^{-1}$ as well as its Jacobian and gradients. Additionally, $T_\theta$ is needed to generate new samples $\bm{x}$ from latent samples $\bm{z} \sim \pi(\bm{z})$. Various architectures of normalizing flows, such as \textit{autoregressive flows}, enable such operations, as reviewed in \cite{Papamakarios2021}. { We provide here some details, without aiming to be exhaustive, to clarify the design of certain architectures used in the literature.
}

The general schema of \textit{autoregressive flows} is as follows: let $\bm{x}, \bm{z} \in \mathbb{R}^d$, and denote $\bm{x} = (x_1, x_2, \dots, x_d)$ and $\bm{z} = (z_1, z_2, \dots, z_d)$. Then $\forall i$:
\begin{equation}
x_i = T(z_i; c_i) \quad \mathrm{with} \quad c_i = C_i(z_1, \dots, z_{i-1}) \triangleq C_i(z_{<i}),
\end{equation}
{where $T$ knows as the \textit{transformer}\footnote{This term should not be confused with transformer networks used in deep learning architectures based on multi-head attention mechanisms \citep{Vaswani2017}.} is a strictly increasing monotonic transformation, and $C_i$ is known as the \textit{conditioner}.}


{ Among the transformers, the class of affine transformations is simple and widely used.} These are defined as:
\begin{equation}
T(z_i; c_i) = e^{\alpha_i} z_i + \beta_i, \quad \mathrm{with} \quad c_i = (\alpha_i(z_{<i}), \beta_i(z_{<i})).
\label{eq-affine-coupling}
\end{equation}
The invertibility of $T$ is guaranteed by the exponential function, and the Jacobian determinant simplifies to the $e^{\alpha_i}$ factor. Affine transformers are used in works such as \citep{DinhKB14, Papamakarios2017a, DinhSB17, Kingma2018}. 
Furthermore, \textit{spline}-based flows have gained traction, as exemplified in \cite{Crenshaw_2024}, where the \texttt{PZFlow} code is applied to model galaxy photometric redshift posterior distributions.

Concerning the conditioners $C_i$, in principle, they can be any functions (or models) that take $z_{<i}$ as input and output $c_i$. However, practical implementations are driven by computational cost. A particular approach is the \textit{coupling layer}, implemented in \texttt{Real NVP}
as proposed in \citep{DinhKB14, DinhSB17}. 
{ In this context, the structure of the transformation involves a split of the $\bm{z}$ vector coordinates into two parts. The first part, $z_{\leq s}$ with $s=d/2$, is left untransformed (eg. $T=Id$), while the transformation of the second part (eg. affine), $z_{>s}$, depends only on $z_{\leq s}$. Notably, the conditioner functions $\bm{\alpha}$ and $\bm{\beta}$ depend exclusively on $z_{\leq s}$ for both the forward and backward transformations. 
}

This splitting schema, implemented in a so-called \textit{coupling layer}, makes the flow-based model computationally efficient. By using deep neural networks, one can achieve complex $(\bm{\alpha},\bm{\beta})$ functions. Designers of complete architectures can compose different \textit{coupling layers}, alternating the order of $\bm{z}$ elements between layers to ensure that all elements are transformed by the end of the sequence of operations. A generalization of this permutation can be achieved using a $1\times 1$ convolution layer.

\begin{figure}
    \centering
    \begin{subfigure}[b]{0.35\columnwidth}
        \centering
        \includegraphics[width=\linewidth]{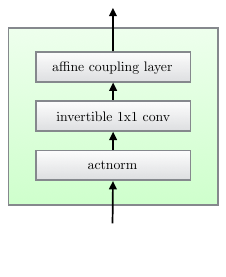}
        \caption{One step of the Glow flow.}
    \end{subfigure}%
    \quad
    \begin{subfigure}[b]{0.5\columnwidth}
        \centering
        \includegraphics[width=\linewidth]{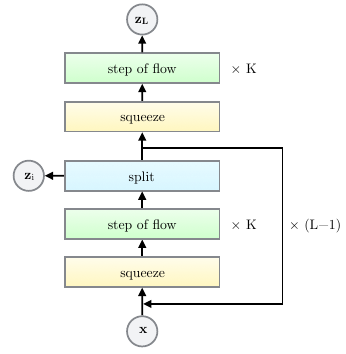}
        \caption{Multi-scale architecture \citep{DinhSB17}.}
    \end{subfigure}
    \caption{Copy of Figure 2 of \citep{Kingma2018}. The flow step consists of a scale and bias layer with data-dependent initialization (\texttt{actnorm}), a $1\times 1$ invertible convolution layer, and an affine coupling layer where $(\bm{\alpha},\bm{\beta})$ are derived from a shallow convolutional neural network. Both $\bm{x}$ and $\bm{y}$ are tensors of shape $[h \times w \times c]$, where $(h, w)$ are the spatial dimensions and $c$ is the channel dimension. The indices $(i, j)$ refer to spatial positions in tensors $\bm{x}$ and $\bm{y}$. This flow step is embedded within a multi-scale architecture consisting of $K$ flow steps and $L$ levels \citep{DinhSB17}. The squeezing operation transforms a tensor of shape $s \times s \times c$ into one of shape $s/2 \times s/2 \times 4c$, mixing spatial and channel components. At each scale, the channel image is divided into $2\times 2$ patches ($s=2$), and each patch undergoes the squeezing operation.}
    \label{fig-Glow-archi}
\end{figure}

\cite{DinhSB17} further improved the modeling by introducing a \textit{multi-scale} approach, which is too extensive to detail here. \cite{Kingma2018} utilized these extensions to build the \texttt{Glow} model, which is employed in our experiments. For completeness, the schema of this model is depicted in Figure~\ref{fig-Glow-archi}.
In the \texttt{Glow} model, the prior is learned during the optimization of the flows, but to generate new samples, one still uses $\bm{z} \sim \mathcal{N}(\bm{0}, \bm{1})$, where $\bm{z}$ comprises tensors corresponding to the number of levels $L$.

\subsection{Score-based diffusion models}
\label{sec-diff}
To generate samples from $p(\bm{x})$, we can proceed as follows \citep[e.g.,][]{Chang2023,LinYang2023}. Starting from a sample $\bm{x}_0$ assumed to be drawn from the distribution $p_0(\bm{x})=p(\bm{x})$, we progressively transform it until the distribution is as simple as, for instance, $\mathcal{N}(\bm{0},\bm{1})$. The idea is then to reverse this process (the \textit{backward} mode) by generating a new sample from a new instance of noise. This transfer mapping is conceptually similar to normalizing flows. Their latent spaces share the same dimension as the data space, avoiding compression as seen in VAE and GAN architectures. However, the main distinction lies in normalizing flows being deterministic by nature, while \textit{score-based diffusion} models are inherently stochastic.

{Using the presentation framework of \cite{guth2022wavelet}}, the transport process in the \textit{score-based diffusion} algorithm is modeled by the Ornstein-Uhlenbeck equation \citep{Uhlenbeck1930}. The \textit{forward} process follows this stochastic differential equation:
\begin{equation}
d\bm{x}_t = -\bm{x}_t \, dt + \sqrt{2}\, dB_t \quad t\in[0,T],
\label{eq-Ornstein-cont}
\end{equation}
where $dB_t$ represents a Wiener process (Brownian motion) and $T$ is a finite endpoint assumed to be sufficiently large. The solution to this forward process is:
\begin{equation}
\bm{x}_t = \bm{x}_0 e^{-t} + \underbrace{(1-e^{-2t})^{1/2}}_{\sigma_t} \bm{z}, \qquad \bm{z} \overset{\text{iid}}{\sim} \mathcal{N}(\bm{0},\bm{1}).
\end{equation}
The corresponding probability density function is expressed as:
\begin{equation}
p_t(\bm{x}_t) = \int p_t(\bm{x}_t,\bm{x}_0) \, d\bm{x}_0 = \int p_t(\bm{x}_t|\bm{x}_0) p_0(\bm{x}_0) \, dx_0,
\label{eq-smoothing-p0-gauss}
\end{equation}
where, based on the Fokker-Planck equation for the Ornstein-Uhlenbeck process:
\begin{equation}
p_t(\bm{x}_t|\bm{x}_0) = \frac{1}{(2\pi \sigma_t^2)^{d/2}} \exp\left\{-\frac{\|\bm{x}_t-\bm{x}_0 e^{-t}\|^2}{2\sigma_t^2}\right\}.
\end{equation}
Thus, $p_t(\bm{x}_t)$ is the convolution of the initial distribution $p_0(\bm{x}_0)$ with a Gaussian of variance $\sigma_t^2$. This gradually blurs the distribution, leading $p_t(\bm{x}_t)$ to converge towards the normal distribution $\mathcal{N}(\bm{0},\bm{1})$.

In contrast to the VAE architecture, where the transformation from $\bm{x}_0$ to $\bm{z}$ is achieved via an \textit{encoder} implemented by a deep neural network, the above stochastic process simply adds noise to the original signal (e.g., a galaxy image). Similarly, the VAE \textit{decoder} or inverse flows are replaced in this model by a stochastic differential equation (damped Langevin dynamics):
\begin{equation}
d\bm{x}_{T-t} = (\bm{x}_{T-t} + 2 \nabla_x \log p_{T-t}(\bm{x}_{T-t}))\, dt + \sqrt{2}\, dB_t \quad t\in[0,T],
\label{eq-backward-diffusion}
\end{equation}
where $s_t = \nabla_x \log p_{T-t}(\bm{x}_{T-t})$ represents the \textit{score} of the probability density at time $T-t$ for the blurred image $\bm{x}_{T-t}$ under a white noise distribution with variance $\sigma_{T-t}^2$. This backward process constitutes the generative direction, enabling $\bm{x}$ samples to be drawn from $\mathcal{N}(\bm{0},\bm{1})$.

{The central challenge in generation lies in estimating the \textit{score}, the key ingredient of the generation process.} Two main approaches have been proposed: (1) deriving a model for $p_t$ that incorporates regularity patterns and correlations to learn the \textit{score} via \textit{score matching} \citep{hyvarinen2005a}, or (2) employing a \textit{denoising} neural network, as introduced by \cite{Bengio2013}, further developed in \cite{Sohl-Dickstein2015,Ho2020,song2021scorebased,song2021maximum}, and recently examined from a mathematical perspective in \cite{kadkhodaie2024generalization}. The first approach has been utilized in works such as \cite{guth2022wavelet,Lempereur2024} with Gibbs energy parameterization. The second has found applications in astrophysics, including galaxy image generation \citep{smith2021}, generation of 21~cm luminosity temperature maps \citep{Zhao2023}, super-resolution of large-scale cosmic structures \citep{Schanz2023}, and Bayesian posterior sampling for weak lensing mass-mapping problems \citep{Remy2023}.

The connection between a perfect \textit{denoiser} model and the \textit{score} has been explored by numerous authors \cite[e.g.,][]{tweedie1947functions,herbert1956empirical,miyasawa1961empirical} and revisited by \cite{Raphan2011} for Gaussian white noise $\mathcal{N}(0,\sigma^2)$ independent of the noiseless signal $\bm{x}$. Irrespective of the prior on $\bm{x}$, the following theorem holds: the estimator $\tilde{\bm{x}}$ of the signal $\bm{x}$, given the noisy signal $\bm{x}_\sigma = \bm{x} + \bm{z}$ with $\bm{z} \sim \mathcal{N}(0,\sigma^2)$, is:
{
\setlength{\belowdisplayskip}{5pt} \setlength{\belowdisplayshortskip}{5pt}
\setlength{\abovedisplayskip}{5pt} \setlength{\abovedisplayshortskip}{5pt}
\begin{equation}
\tilde{\bm{x}} = \bm{x}_\sigma + \sigma^2 \nabla_x \log p(\bm{x}_\sigma).
\label{thm-score-denoising}
\end{equation}
}
Thus, the integration of a \textit{denoiser} network within a generative model operates as follows. First, the network is trained in a straightforward manner to recover original images from their blurred versions, where the blur is induced by Gaussian white noise. The noise variance is not predetermined or explicitly provided to the network during training but spans the same range (e.g., $[0,1]$) as the image values. During the backward stochastic process at each time step $t$, the \textit{denoiser} is applied to obtain a denoised estimate ($\tilde{\bm{x}}$) of the image $\bm{x}_{T-t}$ ($\bm{x}_\sigma$), which has been blurred by white noise with a known variance $\sigma^2 = \sigma^2_{T-t}$. This process yields the score $s_t$ thanks to Equation~\ref{thm-score-denoising}, enabling progression to the next step in the backward direction.

Discussions comparing GANs and diffusion models can be found in \cite{dhariwal2021diffusion}. We do not delve into the potential integration of flow-based and diffusion-based models \citep[e.g.,][]{zhang2021diffusion,gong2021interpreting} or the use of diffusion models to enhance GAN stability, as introduced by \cite{Wang2023}. These topics present intriguing opportunities for exploration in future studies.
%

{
\subsection{Some comments}
\label{sec-some-comments}

In the previous sections, we have provided some background knowledge for practitioners who may find it perplexing to choose a generative model for their specific application. The field is evolving rapidly, making this review inherently challenging. However, we can highlight some key elements that will underpin the following numerical experiments.

Despite the efforts of GAN architecture designers, optimization does not always guarantee convergence to the correct minimum for the generator. This raises concerns about whether $p_G = p_{\text{data}}$, potentially leading to generation failures such as \textit{mode collapse}. Nevertheless, GANs have paved the way for the first generation of deep generative models, exposing fundamental challenges and demonstrating the feasibility of leveraging deep networks in this field.  

While the loss function of normalizing flows aims to minimize an appropriate metric compared to GAN optimization, the challenge of determinant computation still constrains the expressiveness of these models. Subsequently, denoising diffusion stochastic models have gained popularity due to their strong performance, rivaling that of GANs while offering distinct advantages. Notably, these models do not rely on adversarial training, making them simpler to optimize. They are grounded in well-defined mathematical principles, ensuring a theoretically sound training process. However, they depend on an iterative generation scheme, which can be computationally intensive and slower compared to GANs.  

Returning to the original GAN problem, the question \textit{"Does $p_G = p_{\text{data}}$?"} remains an open challenge. This is why numerical experiments are still a valuable approach. We have selected three models, each representing one of the three types described in the previous sections, to assess the reliability of model optimization. This choice may be questioned but we think that it serves more as guide lines if one would like to use other models. 
Regarding the number of model parameters, we maintain them in the range of $O(10)$ million, which is significantly lower than the scale used in large diffusion models mentioned in the introduction. This choice is relevant for two reasons: first, it aligns with the constraints faced by practitioners with limited computational resources; second, it is well-suited to the sizes of the images ($d \times d$) and the dataset ($N$) used in our experiments.  
Anticipating one of the results, we observe that, as a general rule of thumb, the ratio of the total number of available training pixels to the total number of model parameters should be at least on the order of $O(10^2)$ to ensure effective learning.
}

\begin{figure*}
    \centering
        \includegraphics[width=0.8\linewidth]{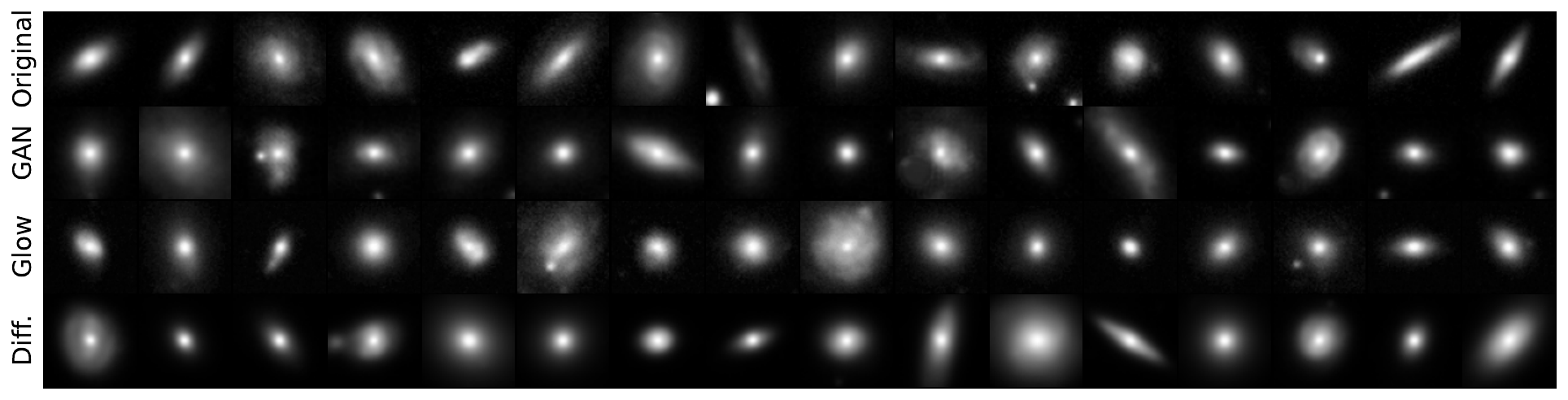}
    \caption{Examples of \textit{validation} images from the original dataset (top row) and generated samples from different models trained with $N = 10^5$ images: the \texttt{light-weight-gan} GAN-based model (second row), the \texttt{Glow} flow-based model (third row), and the diffusion-based model with a \texttt{U-Net} denoiser (bottom row).}
    \label{fig-Original-Glow-UNet-Gan-samples}
\end{figure*}

\section{Experiment}
\label{sec-experiment}
In the previous section, we briefly outlined different architectures of generative models. Here, we describe the process of generating a sample $\bm{x}$, which typically\footnote{This is the usual choice, but alternatives such as uniform distributions can also be considered; see \cite{Brock2019} for a discussion in the context of GANs.} begins with sampling from a centered isotropic Gaussian distribution $\bm{z} \sim \mathcal{N}(\bm{0}, \bm{1})$ (i.e., white noise) of dimension $d_\ell$. { Notably, unlike GANs and VAEs, where latent space compression occurs ($d_\ell \ll d$), diffusion and flow-based models preserve the data dimensionality ($d = d_\ell$).}

An interesting question arises: if we train two generative models of the same architecture and feed them with identical latent variables sampled from the same Gaussian distribution, will the generated images be identical? This could occur if the models have learned the same transformation process to map $\pi(\bm{z})$ to $p(\bm{x})$, or if they simply replicate images from the training dataset, which is not the intended purpose. 

If the generated images differ, how can we evaluate the fidelity of the learning process? More critically, are we truly sampling from the probability density $p(\bm{x})$, or are the generated samples a mixture of the original dataset images?

A partial answer to these questions is provided in \cite{kadkhodaie2024generalization} within the context of diffusion-based models, highlighting the impact of dataset size on the transition from \textit{memorization} to \textit{generation} regimes. Following their methodology, we compare two generative models of the same architecture trained on non-overlapping subsets of the same dataset. For simplicity, we limit our experiment to three models and a single dataset, which we believe suffices to derive meaningful insights. 
{We focus on investigating how one can perform tests to ensure that the model behaves as intended. 
}

%

%
\subsection{The models}
\label{sec-Exp-Models}
We selected three architectures, as described in Section~\ref{sec-generative-models}: a GAN, a flow-based model, and a diffusion-based model. We have used implementations of architectures written in \texttt{PyTorch} \citep{PyTorch2019}. 

\begin{figure}
    \centering
		\includegraphics[width=\linewidth]{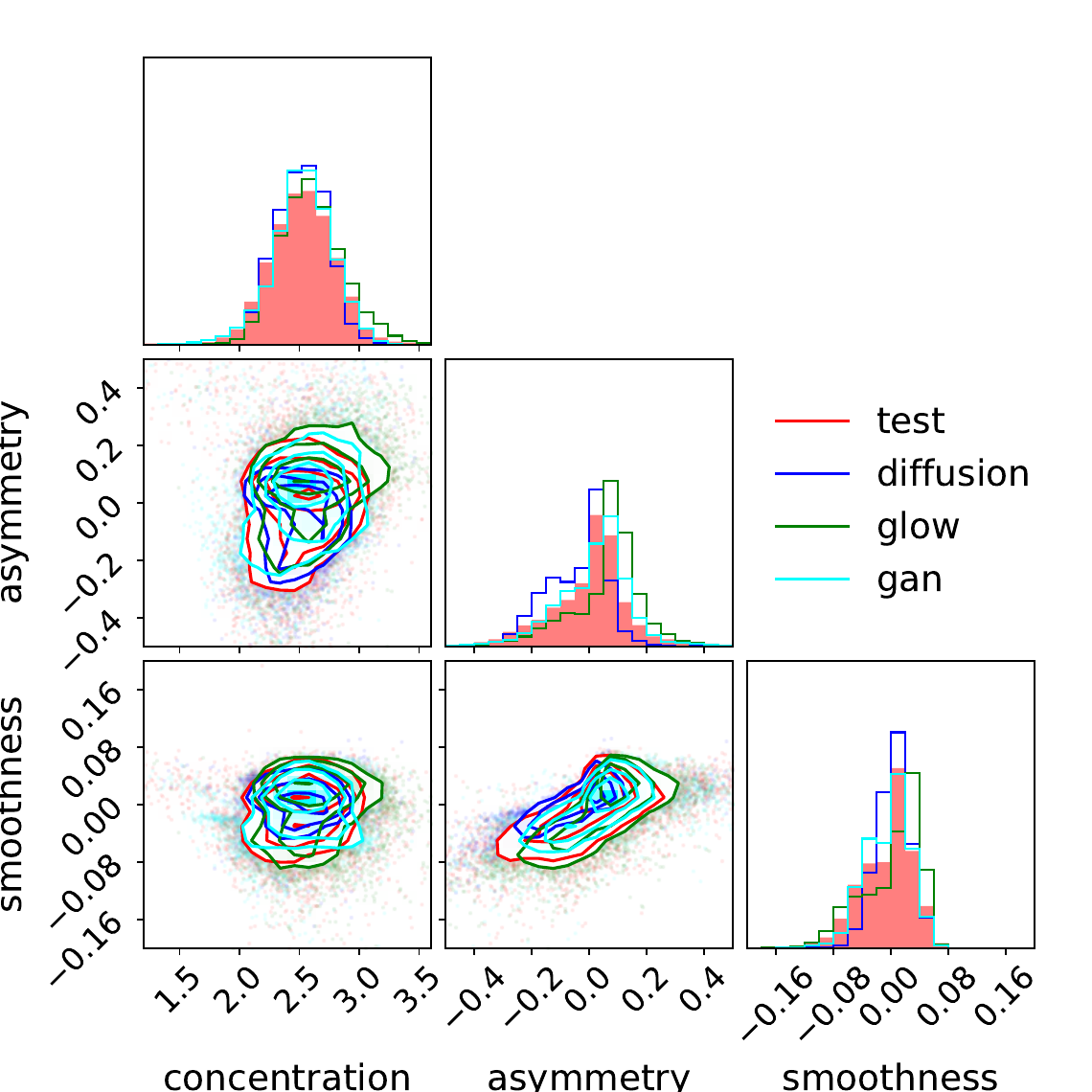}\\
		\includegraphics[width=0.7\linewidth]{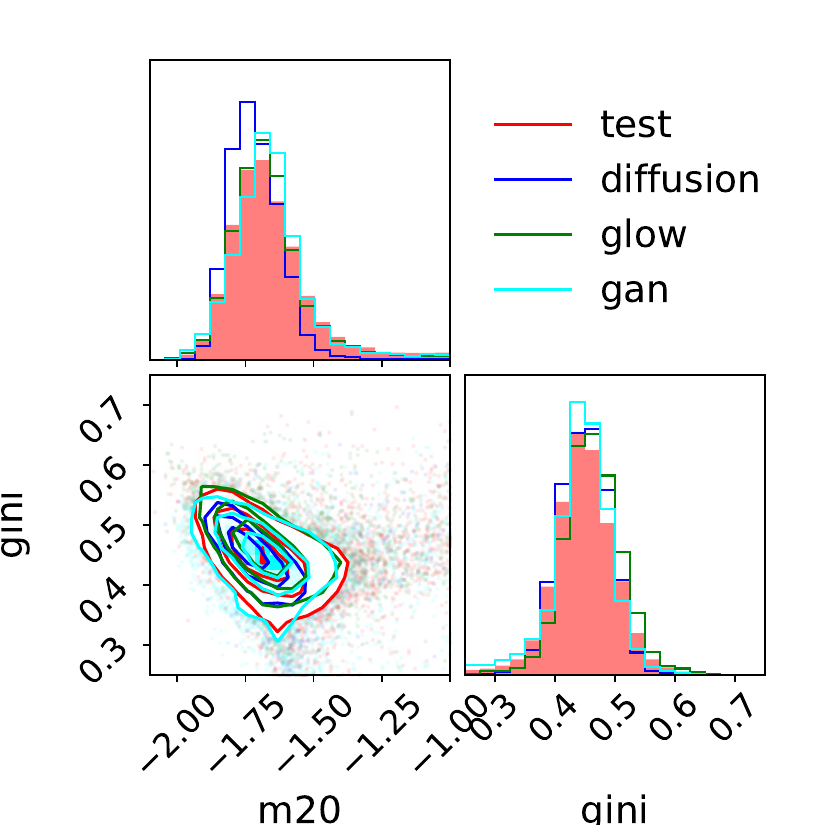}
	\caption{Corner plots\protect\footnotemark of morphological coefficients computed using either the validation dataset unused by any model trainings (\textit{red, "test"}) or generated images from the diffusion-based model (\textit{blue, "diffusion"}), the \texttt{Glow} flow-based model (\textit{green, "glow"}), and the \texttt{light-weight-gan} GAN-based model (\textit{cyan, "gan"}). All models were trained with $N = 10^5$ galaxy images and correspond to the results shown in Figure~\ref{fig-Original-Glow-UNet-Gan-samples}. The plot design is inspired by \protect\cite{HACKSTEIN2023100685}.}
	\label{fig-morpho-coeff}
\end{figure}
\footnotetext{Corner plots generated using the library \url{https://corner.readthedocs.io} \citep{corner2016}.}

For the GAN model, we used the \texttt{light-weight-gan} architecture\footnote{Source code: \url{https://github.com/lucidrains/lightweight-gan/} described in Section \ref{sec-GAN}. Original repository: \url{https://github.com/odegeasslbc/FastGAN-pytorch/tree/main}.}. Default settings were applied, including a latent space dimension of $d_\ell=256$ leading to approximately $36$ million parameters. Training involved resizing input images to $128 \times 128$ pixels and applying horizontal/vertical flips for data augmentation. The training process lasted about $30$ hours over $100,000$ iterations with a batch size of $10$ images.

For the flow-based model, we used the \texttt{Glow} architecture\footnote{Source code: \url{https://github.com/rosinality/glow-pytorch}. Original repository: \url{https://github.com/openai/glow}.} described in Section \ref{sec-NF}. The model configuration, shown in Figure \ref{fig-Glow-archi}, includes $K=32$ flows and $L=3$ blocks or levels by default\footnote{Different $(K,L)$ values have been used with no significant differences with the default values.}, comprising $44$ million parameters. Training followed default settings, with image pixel values reduced from 8-bit to 5-bit precision \citep{Kingma2018}. Optimization had required approximately $24$ hours for about $200,000$ iterations with a batch size of $32$ images.

For the diffusion-based model, we adopted a \texttt{U-Net} architecture \citep{ronneberger2015u} as the \textit{denoiser}, leveraging its multi-scale structure with $7.6$ million parameters\footnote{Source code: \url{https://github.com/LabForComputationalVision/memorization_generalization_in_diffusion_models}.}. Training followed the procedure in \cite{kadkhodaie2024generalization}, with $100$ epochs taking about $100$ hours.

{We have perform some tests with other settings of the above architectures and even if not exhaustive, the results presented in the following section are generic and would serve as guide line if one would like to use either another settings of the models we have used, either to test a completely different model architecture.

All experiments have been conducted mainly on a single Nvidia V100-32g GPU at Jean Zay supercomputer at IDRIS\footnote{\url{http://www.idris.fr/eng/jean-zay/jean-zay-presentation-eng.html}}, a national computing centre for the France's National Centre for Scientific Research (CNRS).
}

\subsection{The dataset}
\label{sec-Dataset}
%
%

{
In the present study, we focus on galaxy image generation, taking as a reference the experiment conducted by \cite{kadkhodaie2024generalization}, who, however, used widely known machine learning datasets as face or bedroom images. For completeness, all their images were grayscale (single channel), with intensity values in the range $[0,1]$ and a resolution mostly of $80 \times 80$ pixels.

We use the Sloan Digital Sky Survey (SDSS) dataset \citep{sdssdr7} as described in \citep{smith2021}\footnote{The script to download the original selected dataset is available at \url{https://github.com/Smith42/astroddpm/data/sdss}.}. As the authors mentioned, the volume complete sample includes $(g,r,z)$ bands, with an $r$-band absolute magnitude limit of $M_r\leq 20$ and a redshift limit of $z\leq 0.08$. Each band image consists of $512\times 512$ pixels at a resolution of $0.262^{\prime\prime} \mathrm{pixel}^{-1}$, with an average seeing of approximately $1^{\prime\prime}$. Our preprocessing applied to the downloaded dataset, consisting of approximately $300,000$ images, includes cropping around the target galaxy coordinates to $64 \times 64$ pixels to ensure that each image predominantly contains a single object. Additionally, a single grayscale channel was created from the $(z, g, r)$ bands using the \texttt{scipy} \citep{2020SciPy-NMeth} library to balance the contributions of the different bands\footnote{The \texttt{make\_lupton\_rgb} function was applied with settings $Q=8$ and \texttt{stretch}=0.2.}. Of course, this choice of dataset is by no means restrictive, and researchers may use their own datasets. To facilitate reproducibility, we provide additional materials in the companion GitHub repository, enabling users to replicate the experiments.

Thus, we restrict ourselves to relatively modest image sizes compared to those produced by models such as \texttt{light-weight-gan} and \texttt{Glow}. Our primary objective is to investigate whether the transition from \textit{memorization} to \textit{generalization} in diffusion models occurs around $10^5$ training images, as mentioned earlier (see Section~\ref{sec-some-comments}). Furthermore, we aim to examine if similar transition occurs for  GANs and flow-based models. 

However, even when using similar image sizes, it remains an open question whether the same number of training images is required to achieve the generalization transition when working with face or bedroom images, which may be more complex than galaxy images at the same resolution. The complexity of such images can be estimated based on the compression ratio achieved using the same algorithm\footnote{For example, \texttt{JPEG} compression with default parameters in the \texttt{PIL} library v11.1.0 (\url{https://pillow.readthedocs.io/}).}. We find that the galaxy images used in this study exhibit a higher compression ratio, approximately $50\%$ greater. Thus, for less complex images, one might expect generative models to have an easier task. At the very least, this question can be explored within the framework described here.
}

\subsection{Results}
\label{sec-Results}
\subsubsection{Generated images}
\label{sec-Generated-Images}
Figure~\ref{fig-Original-Glow-UNet-Gan-samples} shows examples of galaxy images (\textit{validation}) from the original dataset alongside samples generated by the three different models trained with $N = 10^5$ images.
At first glance, the generated images appear indistinguishable from the original images. To provide a more quantitative evaluation, Figure~\ref{fig-morpho-coeff} presents various morphological coefficients computed using \texttt{Statmorph}\footnote{\url{https://github.com/vrodgom/statmorph}} \citep{2019MNRAS.483.4140R}, including Concentration, Asymmetry, and Smoothness \citep{2000AJ....119.2645B,2003ApJS..147....1C,2004AJ....128..163L}, as well as Gini and $M_{20}$ coefficients \citep{2004AJ....128..163L,10.1093/mnras/stv2078}.

The distributions of morphological coefficients derived from the three models show good agreement with those computed from the original images. However, we do not aim for high-quality images or state-of-the-art results, as seen in \citep{ravanbakhsh2016, Fussell2019, Lanusse2021, smith2021, HACKSTEIN2023100685}, which achieve better image quality. Instead, our focus is on a different question: with low-resolution and relatively simple images, do models of the same type trained on moderately sized datasets learn the same underlying probability distribution? Furthermore, if differences exist, can we identify measures to assess the reliability of the results?
\subsubsection{\texttt{U-Net} Denoising Performances}
\label{sec-UNet-Perf}
%
{ Before assessing the generative performance of the diffusion models in the next section, we first present the denoising performance of two \texttt{U-Net} networks with the same architecture, which are subsequently used to estimate the score as discussed in Section~\ref{sec-diff} (Equations~\ref{eq-backward-diffusion},\ref{thm-score-denoising}).
}

\begin{figure}
    \centering
    \includegraphics[width=0.95\linewidth]{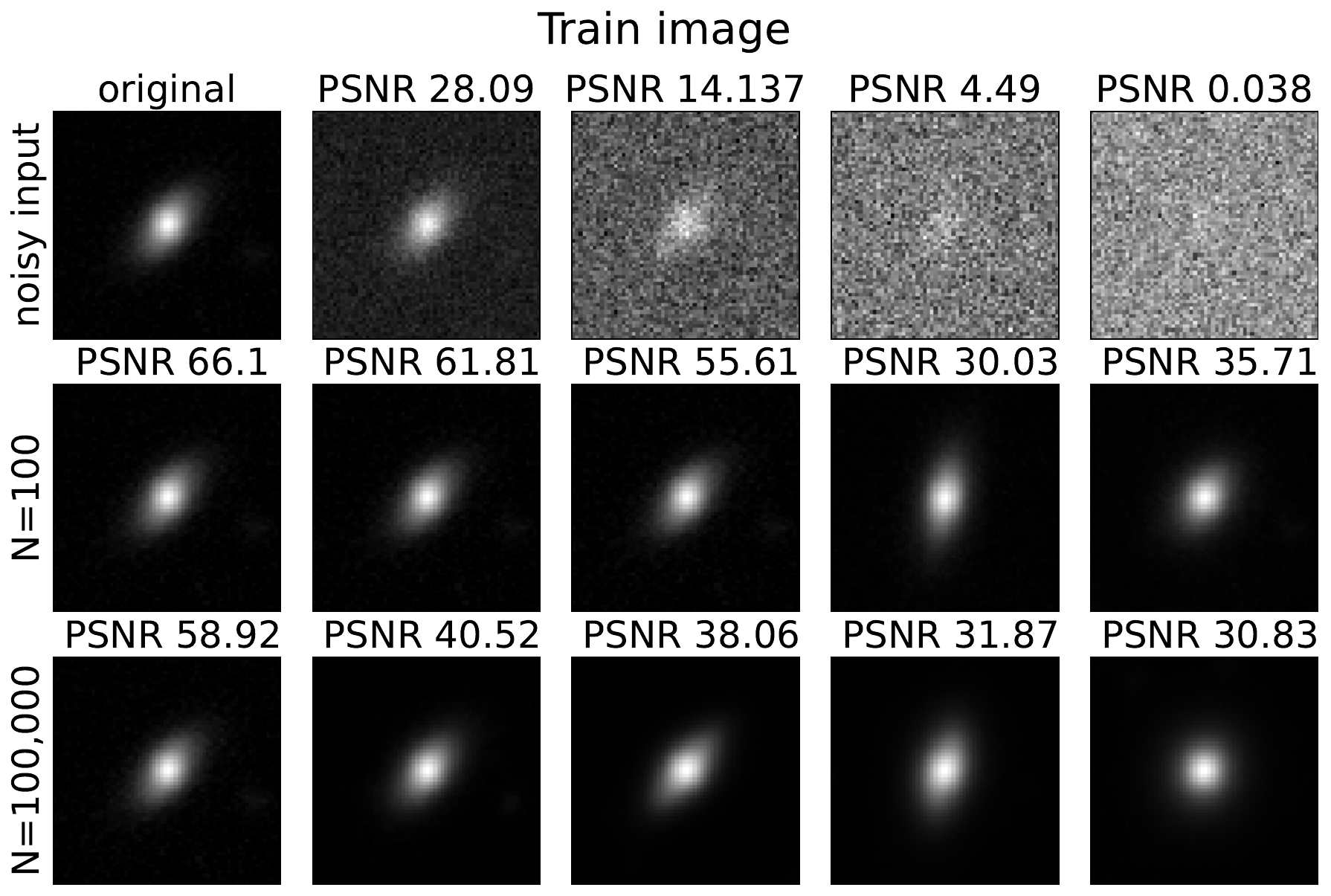}
    \caption{Denoising performances of two \texttt{U-Net} models on an image from the \textit{training sample}. 
    Top row: (left to right) the original image with progressively increasing noise added. The \texttt{PSNR} is defined as $10\log_{10}$ ratio of the squared dynamic range to the mean square error. 
    Middle row: (left to right) denoised images and their PSNR values obtained using a \texttt{U-Net} trained with a dataset of size $N=100$. 
    Bottom row: same as the middle row but with a \texttt{U-Net} trained with a much larger dataset of $N=100,000$ images.}
    \label{fig-UNet-denoising-train}
\end{figure}
\begin{figure}
    \centering
    \includegraphics[width=0.95\linewidth]{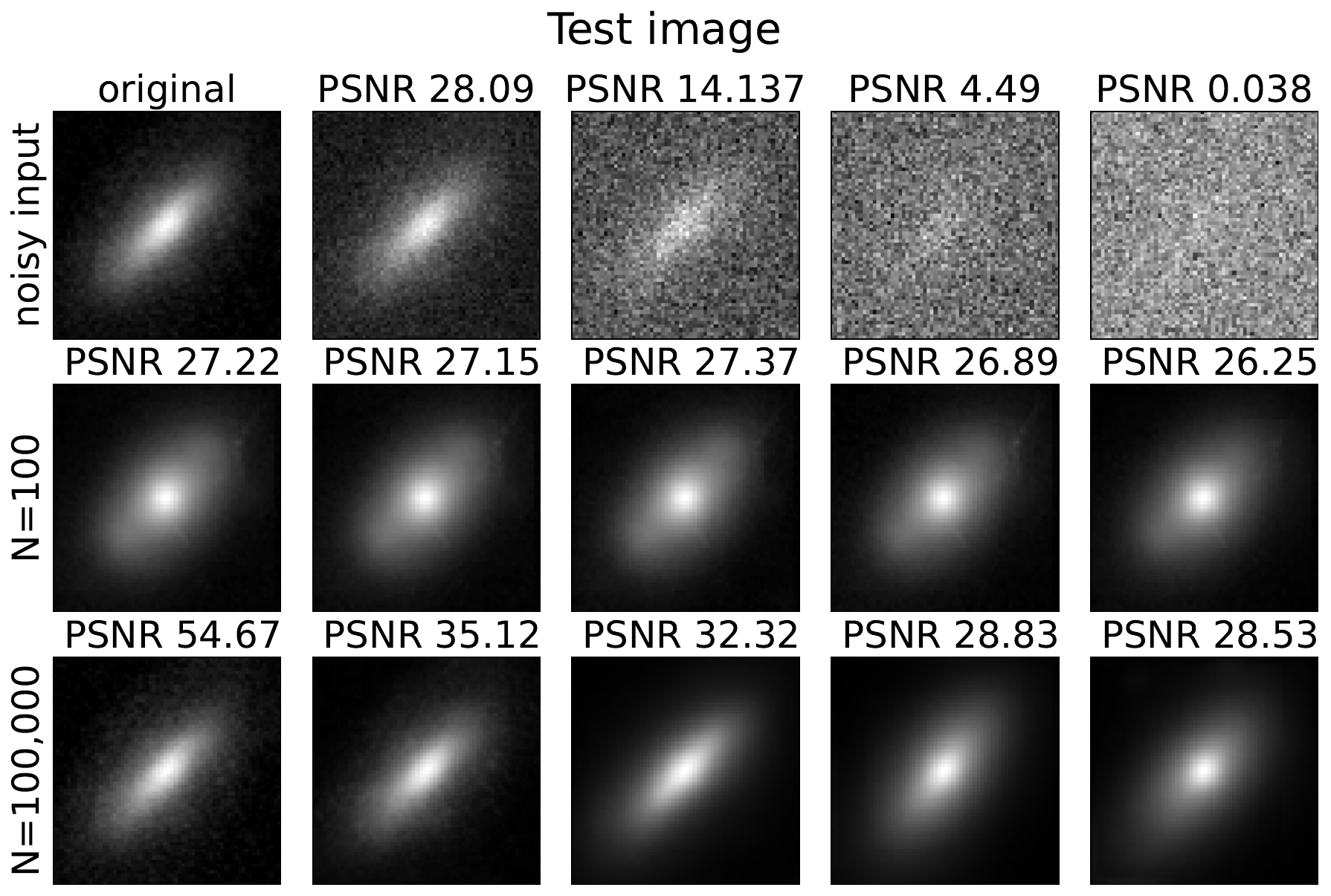}
    \caption{Similar to Figure~\ref{fig-UNet-denoising-train}, but applied to an image that was never used to train the networks (a \textit{validation} image).}
    \label{fig-UNet-denoising-test}
\end{figure}

Figure~\ref{fig-UNet-denoising-train} presents the denoising results of the networks trained on datasets of different sizes, using an original image shared by both training sets. In contrast, Figure~\ref{fig-UNet-denoising-test} demonstrates the performance on an input image that was never included in the training data of either network, referred to as a \textit{validation image}. For both figures, the peak signal-to-noise ratio (PSNR) relative to the original image is provided above each image. { This metric is appropriate, as it directly relates to the loss function used to optimize the denoiser. Additionally, in the context of score estimation, we focus on the overall denoising performance across the entire image, not just the galaxy pixels.} The top row shows the input images (original or noisy) fed to the denoiser. The middle row displays the outputs from the \texttt{U-Net} trained with $N=100$ images, while the bottom row shows outputs from the \texttt{U-Net} trained with $N=100,000$ images. It is important to note that the networks were trained using noisy images without access to PSNR values or noise variance information.

When using an image from the \textit{training} dataset, both networks demonstrate similar denoising capabilities, as shown by the evolution of the output images' PSNR values relative to the PSNR of the noisy input images. However, a notable difference emerges when using the \textit{validation image}. The \texttt{U-Net} trained with $N=100,000$ images achieves results comparable to those observed with the training image, indicating good generalization performance. In contrast, the \texttt{U-Net} trained with $N=100$ images shows limited generalization: the PSNR values of its output images remain nearly constant ($\mathrm{PSNR} \approx 27$), even for input images with very low noise levels. 

{This disparity in denoising performance between the two networks may have significant implications for the generative process, as the score computation could be problematic, starting with a purely noisy image and progressively refining it into a noiseless output.}
\subsubsection{Two Diffusion-Based Models Test}
\label{sec-two_models-Diffusion}
\begin{figure*}
    \centering
        \includegraphics[width=0.9\textwidth]{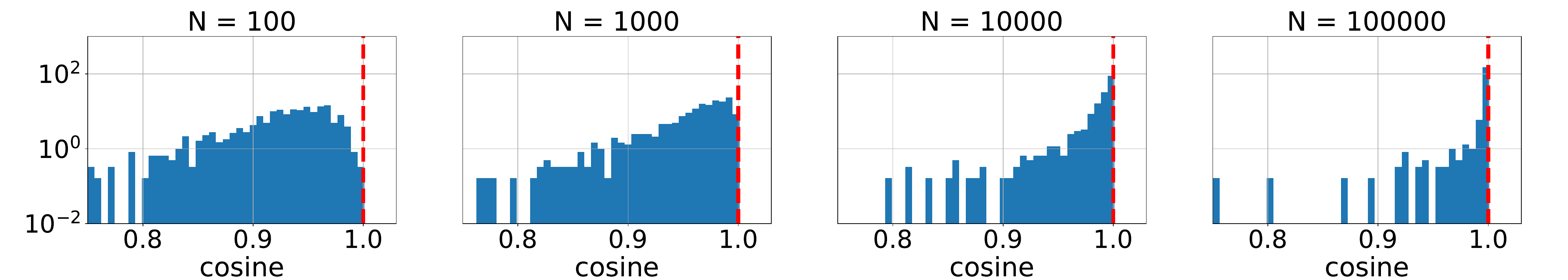}\\
    \includegraphics[width=0.9\textwidth]{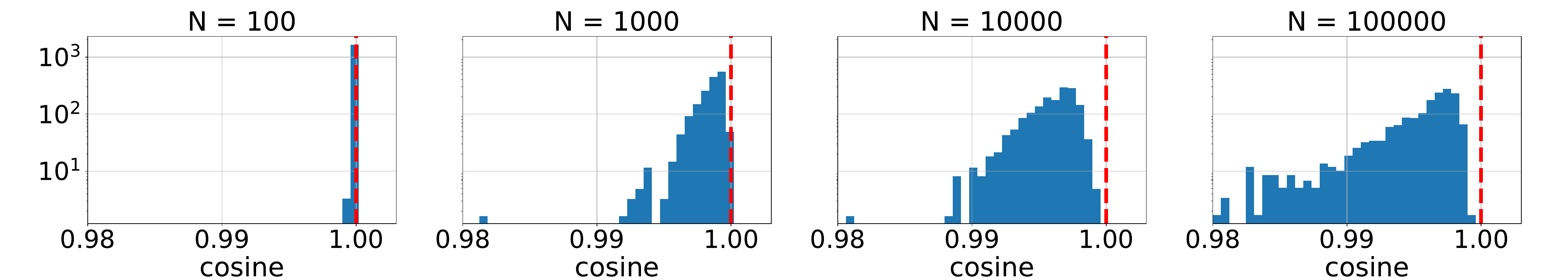}
    \caption{Evolution of the cosine similarity metric for diffusion-based models. 
    Top row: normalized histograms of cosine similarity values between $1,000$ samples generated by two models trained on independent datasets of size $N$, using the same white noise images as seeds. 
    Bottom row: maximum cosine similarity between $1,000$ generated samples from a single model and all images in its corresponding training dataset (i.e., the cosine of the closest training image for each generated sample). 
    The logarithmic vertical scale and linear horizontal scale (cosine value) are consistent across all histograms on each row. The horizontal range in the bottom row is restricted to values near the maximum (cosine = 1), indicated by the dashed red vertical line. This figure is inspired by Figure 2 of \protect\cite{kadkhodaie2024generalization}.}
    \label{fig-cosine-diffusion}
\end{figure*}

Figure~\ref{fig-cosine-diffusion} explores the evolution of the cosine similarity metric \citep{books/aw/TanSK2005}, considering two distinct scenarios. The first scenario (top plots) involves $1,000$ samples generated by two diffusion-based models ("Model A" and "Model B") trained on independent datasets of the same size $N$, with generation \textit{seeded by the same white noise images}. The second scenario (bottom plots) examines a single model, comparing each of its $1,000$ generated samples to the closest training image in terms of cosine similarity.

Although the training images are less complex than datasets like \texttt{CelebA} (faces) or \texttt{LSUN} (bedrooms), the following observations emerge:
\begin{itemize}
    \item {Low dataset size ($N=100$):} 
    The generated samples are highly similar to the training images (cosine similarity close to 1), as indicated by the bottom-left histogram. However, samples generated from the same white noise image by the two models differ significantly, as the cosine similarity values are far from 1 (upper-left histogram).
    \item {Larger dataset sizes:} 
    As the dataset size increases, generated samples become increasingly distinct from training images (progression from left to right histograms in the bottom row). Simultaneously, samples generated by the two models using the same noise seeds become more similar to each other (progression from left to right histograms in the top row). For $N=100,000$, the two models produce almost identical outputs, though not perfectly (a single bin at cosine similarity of 1 has not yet been achieved).
\end{itemize}
These findings highlight two key behaviors: (i) At small dataset sizes, \texttt{U-Net}-based diffusion generative models operate in a \textit{memorization regime}, where they predominantly replicate the training data; (ii) As the dataset size increases, the models transition into a \textit{generation regime}, enabling them to generalize and produce novel images. These behaviors align with observations reported in \cite{kadkhodaie2024generalization} and are fundamentally tied to the denoising capabilities discussed in the preceding section.

\begin{figure}
    \centering
	\includegraphics[width=0.7\linewidth]{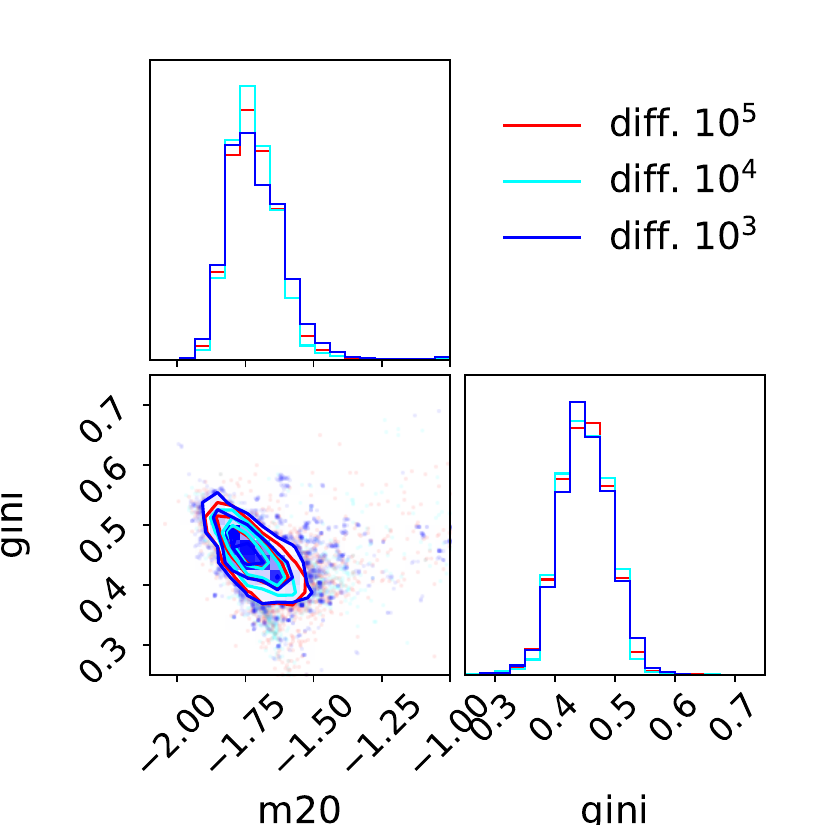}
	\caption{Some morphological coefficients (see Figure~\ref{fig-morpho-coeff}) computed using diffusion models trained with different dataset sizes ($N=10^5, 10^4, 10^3$).}
	\label{fig-morpho-coeff-diff-comparison}
\end{figure}

Nonetheless, an examination of the morphological variable distributions (Figure~\ref{fig-morpho-coeff}) reveals no significant changes when the dataset size is reduced from $10^5$ to $10^3$ (Figure~\ref{fig-morpho-coeff-diff-comparison}). However, reducing the training dataset size leads to generate samples that look similar to training images. This suggests that relying solely on morphological variable is insufficient to capture the profound differences between a model operating in a \textit{memorization regime} and one functioning in a \textit{generation regime}.

{ 
Interestingly, and perhaps counterintuitively, even for relatively simple, low-resolution galaxy images---considered less complex than face or bedroom images, as discussed in Section~\ref{sec-Dataset}---approximately the same amount of $10^5$ samples is required to reliably achieve the transition into the \textit{generation regime}. However, for practical applications, as emphasized in \cite{kadkhodaie2024generalization}, producing higher-resolution images does not require a proportional increase in the number of samples. This is attributed to the hierarchical and multi-scale structure inherent to the images.
}

The "two-models test" conducted with diffusion-based architectures suggests that similar experiments could be useful to scrutinize what is learned by \texttt{Glow} flow-based and \texttt{light-weight-gan} GAN-based architectures.
\subsubsection{Flow-Based Models Tests}
\label{sec-two_models-Flow}
\begin{figure}
    \centering
    \includegraphics[width=\linewidth]{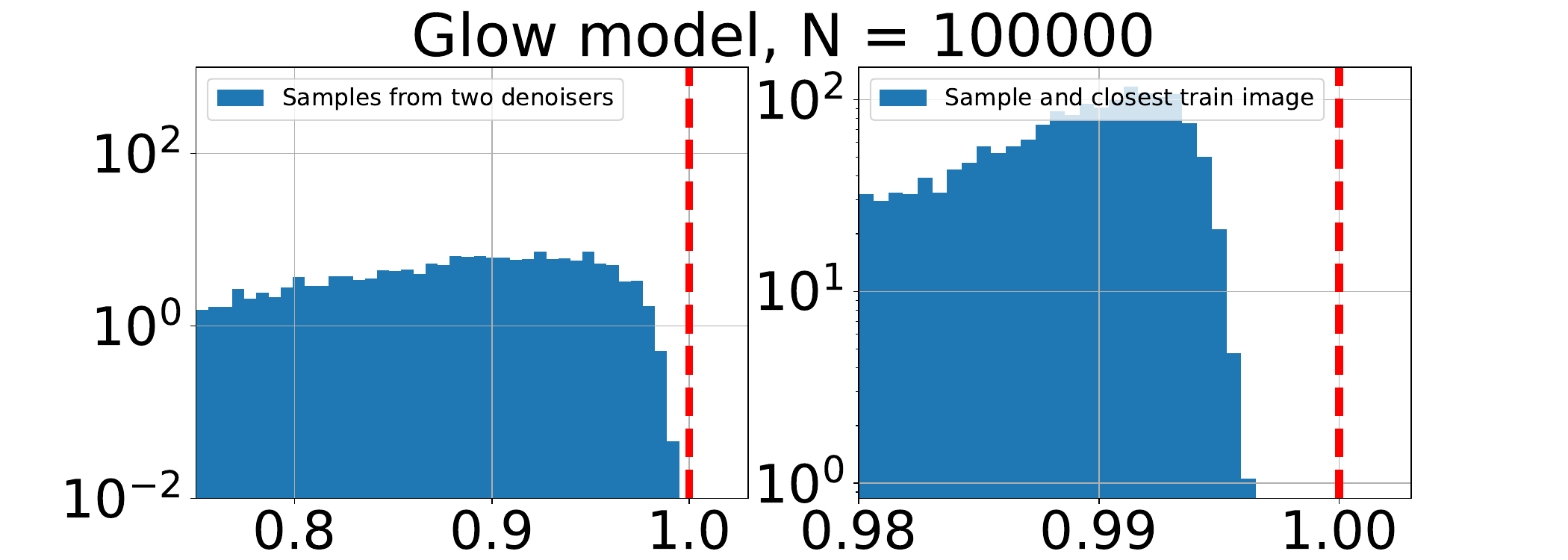}
    \caption{Histograms similar to those of Figure~\ref{fig-cosine-diffusion} for two \texttt{Glow} models trained on independent datasets of size $N=100,000$ (identical to the datasets used by the diffusion models). 
    Left: cosine similarity of $1,000$ samples generated by the two models seeded with the same white noise image. 
    Right: maximum cosine similarity values of $1,000$ samples generated by one model compared to all training images and their augmented versions (horizontal, vertical, and horizontal+vertical flips), as these augmentations are part of the \texttt{Glow} training process.}
    \label{fig-cosine-glow}
\end{figure}

Following the two-models test described in Section~\ref{sec-two_models-Diffusion}, we trained two flow-based models ("A" and "B") with independent galaxy datasets of size $N=10^5$. The cosine similarity results are shown in Figure~\ref{fig-cosine-glow}, { which maintains the same settings\footnote{This results in a cut on the left side of the histograms, but our focus is on the region close to a cosine similarity of around 1.} as Figure~\ref{fig-cosine-diffusion}}. We can draw some key observations from these results:
\begin{itemize}
    \item The generated samples appear distinct from any training image (or their augmented versions) as the right histogram does not peak at a cosine similarity of 1. 
    \item Even when seeded with the same white noise, the images generated by the two models are different, as indicated by the absence of a peak at cosine similarity 1 in the left histogram.
\end{itemize}

This raises a fundamental question: what has been learned by the models?
To further investigate, we utilize the bijective property of flow-based models (Section~\ref{sec-NF}). This "inversion test" proceeds as follows:
\begin{itemize}
    \item Sample $\bm{z} \sim \mathcal{N}(\bm{0}, \bm{1})$.
    \item Use Model "A" in \textit{reverse/backward} mode to generate new samples $\bm{x}_A = G_A(\bm{z})$.
    \item Apply either Model "A" in \textit{forward} mode to get $\bm{z}_A^\prime = G_A^{-1}(\bm{x}_A)$, or similarly Model "B" to get $\bm{z}_B^\prime = G_B^{-1}(\bm{x}_A)$.
\end{itemize}
The central question is whether $\bm{z}_{\texttt{model}}^\prime$ (where "\texttt{model}" stands for "A" or "B") matches $\bm{z}$ in terms of probability density, and more strictly, whether $\bm{z}_{\texttt{model}}^\prime = \bm{z}$ in a one-to-one correspondence.
\begin{figure}
    \centering
    \includegraphics[width=0.7\linewidth]{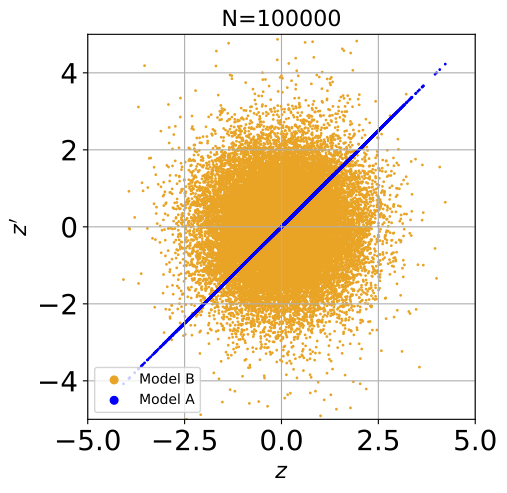}\\
    \includegraphics[width=0.6\linewidth]{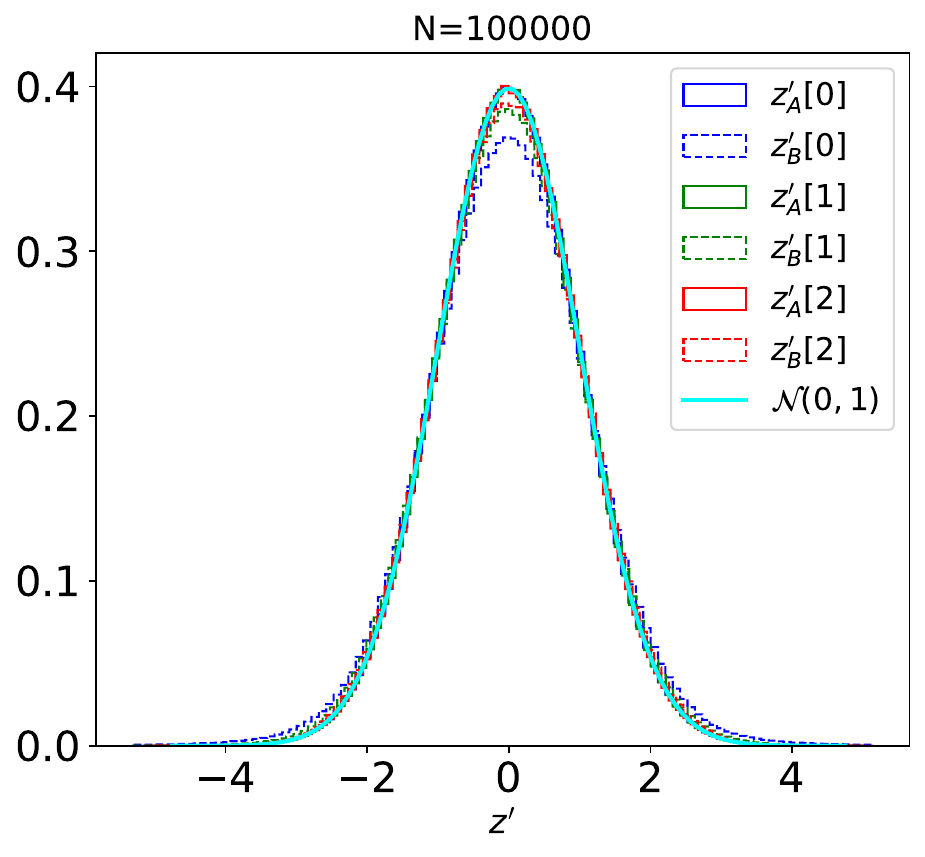}
    \caption{Results of the "inversion test" using $1,000$ samples $\bm{z}$ (see text). 
    Top panel: Comparison of $z^\prime_{\texttt{model}}$ vs. $z$ at all levels of the \texttt{Glow} architecture (see Figure~\ref{fig-Glow-archi}) for Model A (blue dots) and Model B (orange dots). 
    Bottom panel: Distributions of $z^\prime_{\texttt{model}}[\ell]$ for different levels $\ell = \{0, 1, 2 = L\}$, where solid (resp. dashed) curves correspond to Model A (resp. Model B). Both models were trained on independent datasets of $10^5$ galaxy images.}
    \label{fig-glow-inversibility}
\end{figure}
Figure~\ref{fig-glow-inversibility} presents the results of the inversion test:
\begin{itemize}
    \item In the left panel, $\bm{z}_A^\prime = \bm{z}$ (blue dots), confirming the consistency of Model A with the flow property $G_A^{-1} \circ G_A = \mathds{1}$. Conversely, $\bm{z}_B^\prime$ and $\bm{z}$ are independent variables, indicating $G_B^{-1} \circ G_A \neq \mathds{1}$.
    \item The right panel shows that for all levels of the \texttt{Glow} architecture, $\bm{z}_B^\prime$ closely follows a $\mathcal{N}(\bm{0}, \bm{1})$ density distribution. While not perfect, $\bm{z}_B^\prime$ approximately retains the same probability density as $\bm{z}$, which is a satisfactory outcome.
\end{itemize}
These results explain why models "A" and "B" cannot render the same image if they are fed with the same latent $\bm{z}$: they differ by a bijection from $\mathcal{N}(\bm{0}, \bm{1})$ to itself. In fact, flow-based models cannot guarantee more than this property. Fortunately, in this situation, both models account for the same sample probability density. However, there is a caveat: while the distributions of the latent variables match, this does not ensure that the output probability density matches the underlying density of the input dataset. To verify this, it is essential to analyze, for instance, the distributions of morphological variables as shown in Figure~\ref{fig-morpho-coeff}.
\begin{table*}
    \renewcommand{\arraystretch}{1.5}
    \centering
    \caption{Values of the 1-Wasserstein metric ($W_1(X,Y)$) computed for \texttt{Glow} models trained in the "two-models test" context with three different training dataset sizes. For $K=32$ flows and $L=3$ levels, and considering $1,000$ generated images, one obtains $6\times10^6$ (resp. $3\times10^6$) $z$ and $z^\prime$ samples at level 0 (resp. levels 1, 2). Using i.i.d. samples $(X,Y)$ from $\mathcal{N}(0,1)$ yields $W_1(X,Y) \approx 7\times10^{-4}$ (resp. $\approx10^{-3}$) for level 0 (resp. levels 1, 2). The very low value of $W_1(z^\prime_A, z) = O(10^{-6})$ is a direct consequence of the $z^\prime_A = z$ property shown in the left panel of Figure~\ref{fig-glow-inversibility}.}
    \begin{tabular}{|c|cc|cc|cc|}
    \hline
        \multicolumn{1}{c}{} & \multicolumn{2}{c}{$N=10^5$} & \multicolumn{2}{c}{$N=10^4$} & \multicolumn{2}{c}{$N=10^3$} \\ \hline
        \multicolumn{1}{c|}{Level} & $z^\prime_A$ vs $z$ & $z^\prime_B$ vs $z$
                                 & $z^\prime_A$ vs $z$ & $z^\prime_B$ vs $z$
                                 & $z^\prime_A$ vs $z$ & $z^\prime_B$ vs $z$ \\ \hline
        0 & $3.7\times10^{-6}$ & $8.7\times10^{-2}$  & $2.3\times10^{-6}$ & $7.6\times10^{-2}$ & $2.6\times10^{-6}$ & $7.9\times10^{-2}$  \\
        1 & $2.6\times10^{-6}$ & $3.6\times10^{-2}$  & $4.0\times10^{-6}$ & $4.8\times10^{-2}$ & $2.7\times10^{-6}$ & $1.1\times10^{-1}$  \\
        2 & $4.1\times10^{-6}$ & $1.9\times10^{-2}$  & $3.9\times10^{-6}$ & $2.7\times10^{-2}$ & $5.1\times10^{-6}$ & $2.9\times10^{-2}$  \\
        \hline
    \end{tabular}
    \label{tab:Wasserstein-1-Glow-AB}
    \renewcommand{\arraystretch}{1.}
\end{table*}

This raises the question: how does reducing the dataset size affect the results? To quantify this, we computed the 1-Wasserstein distance\footnote{Using the \texttt{wasserstein\_distance} function from the \texttt{scipy} library.}, denoted as $W_1(X,Y)$, for two sets of samples $X = \{x_i\}_i$ and $Y = \{y_j\}_j$. Here, $X$ corresponds to the input $z$ samples, while $Y$ corresponds to the output $z^\prime_A$ or $z^\prime_B$ samples at all levels of the \texttt{Glow} architecture (denoted $W_1(\bm{z}^\prime_A, \bm{z})$ and $W_1(\bm{z}^\prime_B, \bm{z})$). Table~\ref{tab:Wasserstein-1-Glow-AB} summarizes the results for varying dataset sizes ($N=10^3, 10^4, 10^5$).

\begin{figure}
    \centering
	\includegraphics[width=0.6\linewidth]{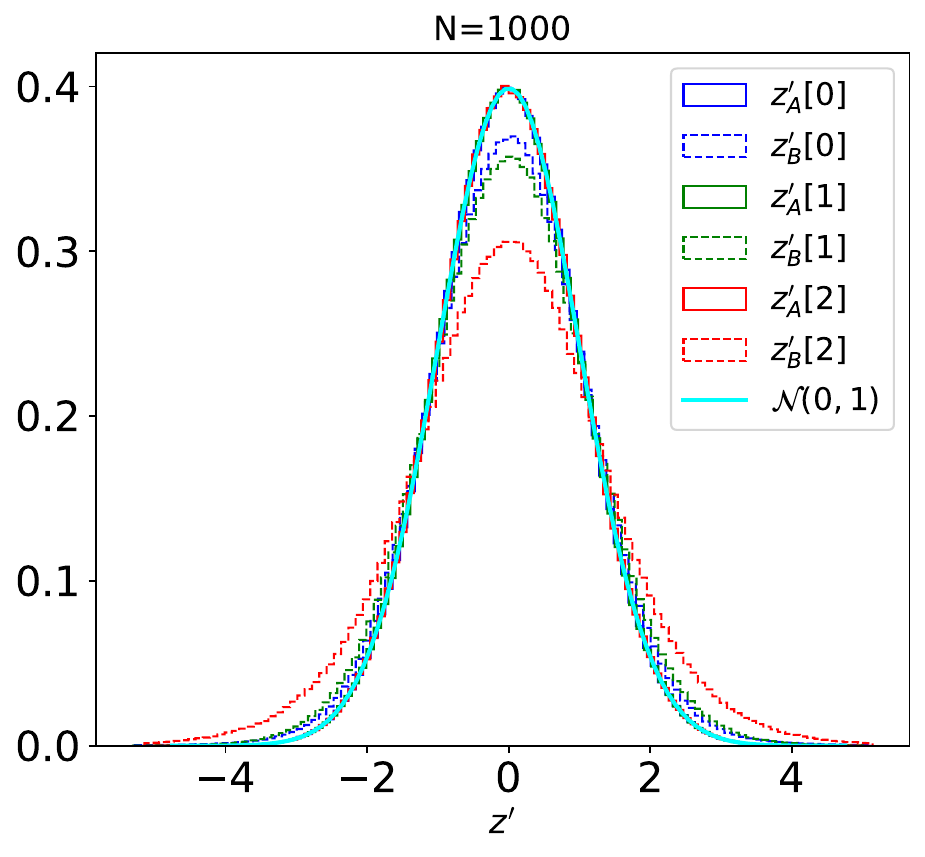}
	\caption{Same legend as bottom panel of Figure \ref{fig-glow-inversibility} but the models have been trained with independent datasets composed of $10^3$ galaxy images. One notices that de $z_B^\prime$ distributions significantly tend to deviate from the $z$ (i.e. $\mathcal{N}(\bm{0},\bm{1})$) contrary to the $z_A^\prime$ distributions.}
	\label{fig-glow-inversibility-1000} 
\end{figure}

\begin{figure}
    \centering
	\includegraphics[width=0.7\linewidth]{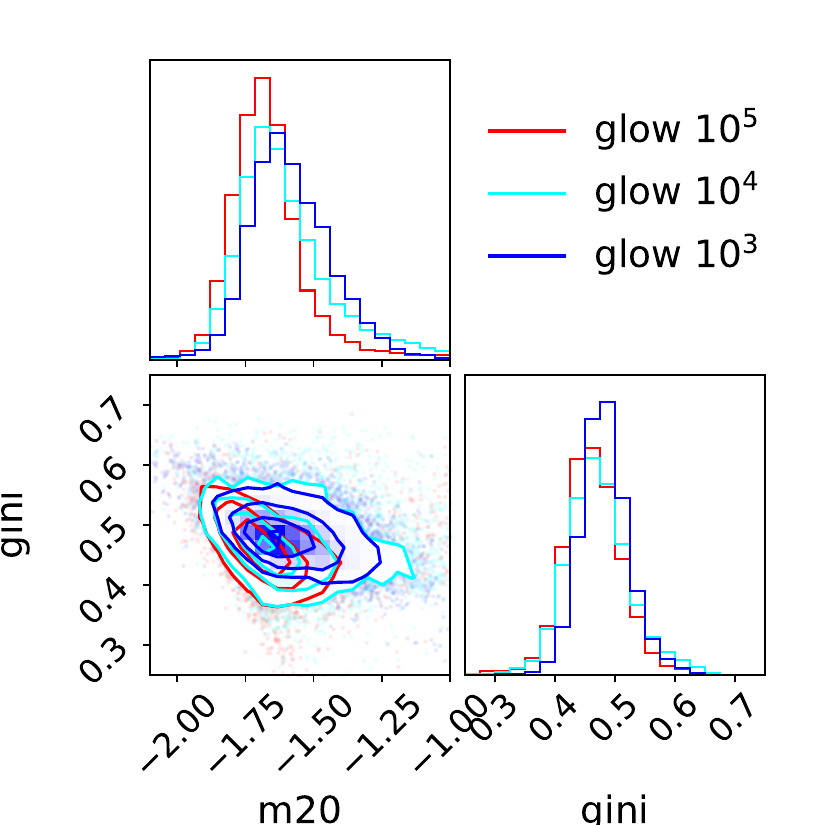}
	\caption{Some morphological coefficients (see Figure~\ref{fig-morpho-coeff}) computed using \texttt{Glow} models trained with different dataset sizes ($N=10^5, 10^4, 10^3$). We can notice a degradation of the $M_{20}$ distribution when the number of training images decreases.}
	\label{fig-morpho-coeff-glow-comparison}
\end{figure}

Decreasing the dataset size significantly degrades the results, as visually evident in Figures~\ref{fig-glow-inversibility-1000} and \ref{fig-morpho-coeff-glow-comparison}. This effect is particularly pronounced for the smallest dataset size, $N=10^3$. However, even with $N=10^4$, the morphological distributions deteriorate, despite the $W_1$ metrics showing relatively small changes compared to the $N=10^5$ dataset results.

\subsubsection{GAN-based models tests}
\label{sec-two_models-GAN}
\begin{figure}
    \centering
	\includegraphics[width=\linewidth]{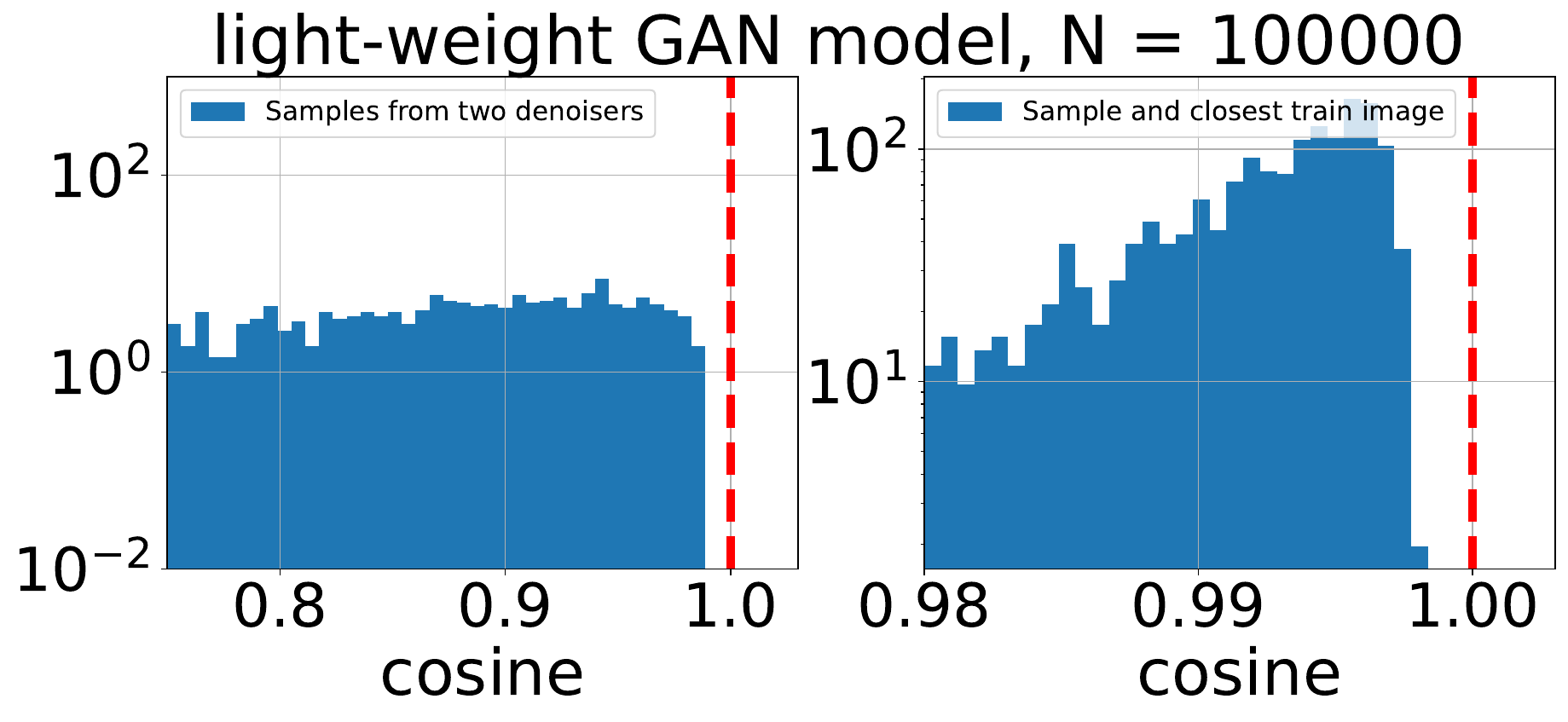}
	\caption{Histograms similar to those of Figures \ref{fig-cosine-diffusion}, \ref{fig-cosine-glow} for two \texttt{light-weight} GAN models using two different datasets of size $N=100,000$ (nb. they are the same as the ones used by the diffusion models). Left: the cosine similarity of $1,000$ samples generated by the two models seeded by the same white noise image. Right: the maximum cosine similarity values of $1,000$ samples generated by one of the models and the training images and their flipped versions (horizontal, vertical and horizontal+vertical) as this data augmentation is used in the training process.}
	\label{fig-cosine-gan} 
\end{figure}

Following the tests done using the diffusion-based models (Section~\ref{sec-two_models-Diffusion}) and the flow-based models (Section~\ref{sec-two_models-Flow}), we trained two GAN-based models ("A" and "B") with independent galaxy datasets of size $N=10^5$. The cosine similarity results are shown in Figure~\ref{fig-cosine-gan}. The results closely resemble those obtained with flow-based models (Figure~\ref{fig-cosine-glow}), exhibiting similar observations: the generated images differ from the training images, and the two models seeded with identical latent variables produce distinct samples. This raises a legitimate question: can we be confident in the model learning?

Although GAN-based models present a different structure compared to flow-based models, as the latent space and dataset space have different dimensionalities, the "inversion test" described in Section~\ref{sec-two_models-Flow} cannot be applied directly. Instead, we propose an alternative approach based on discriminator performance (Section~\ref{sec-GAN}), hereafter referred to as the "discriminator test". According to Theorem 3.1 in \cite{Lim2017}, the hinge loss (Equation~\ref{eq-discri-hinge-loss}) should theoretically converge to a value of 2, provided that both the discriminator and generator are well optimized. This convergence indicates that the probability density of generated samples matches the empirical probability density. To evaluate this, we define the following loss function:
\begin{equation}
    \ell(a,b) = \max(0,1-a) + \max(0,1+b)
    \label{eq-hinge-loss-ab}
\end{equation}
where \(a\) and \(b\) are selected from the set $\{D_M(\bm{x}_M)$, $D_M(\bm{x}_{M^\prime})$, $D_M(\bm{x}_{\text{train}})\}$. Here, $D_M$ represents the discriminator of model $M$ (either "A" or "B"), $\bm{x}_M = G_M(\bm{z})$ are samples generated by the generator $G_M$ using latent variables $\bm{z} \sim \mathcal{N}(\bm{0},\bm{1})$ (identical for both models), and $\bm{x}_{\text{train}}$ refers to training dataset samples. For example, we examine whether a sample $\bm{x}_A$ generated by $G_A$ produces the same discriminator score when replaced by either \(\bm{x}_B\) or $\bm{x}_{\text{train}}$.

\begin{figure}
\centering
\setlength\tabcolsep{0pt}
   \includegraphics[width=0.8\linewidth]{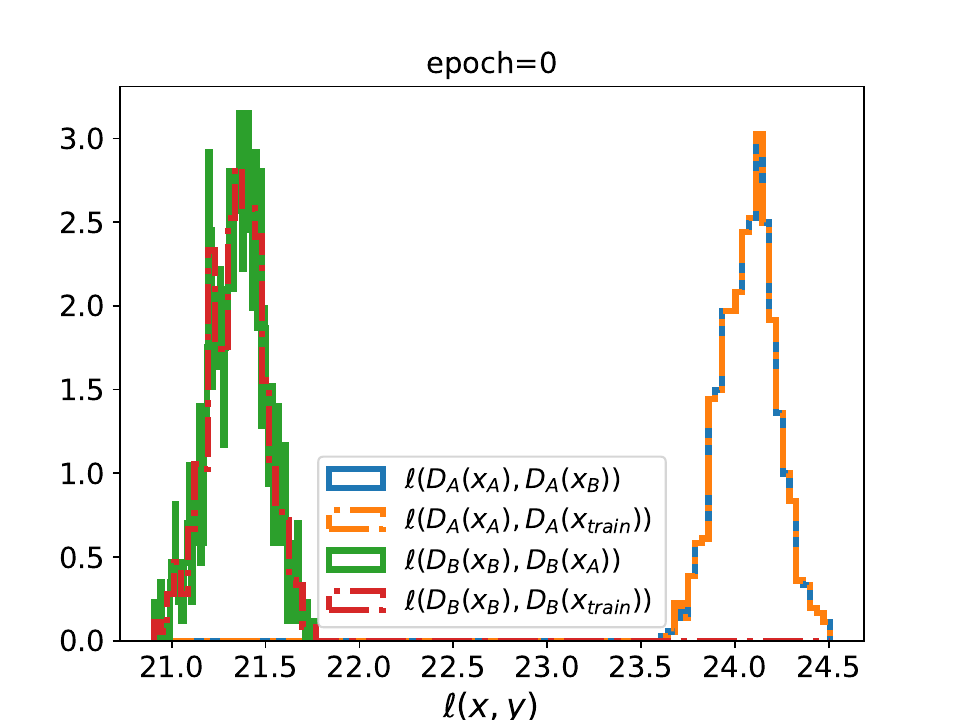}
    \includegraphics[width=0.8\linewidth]{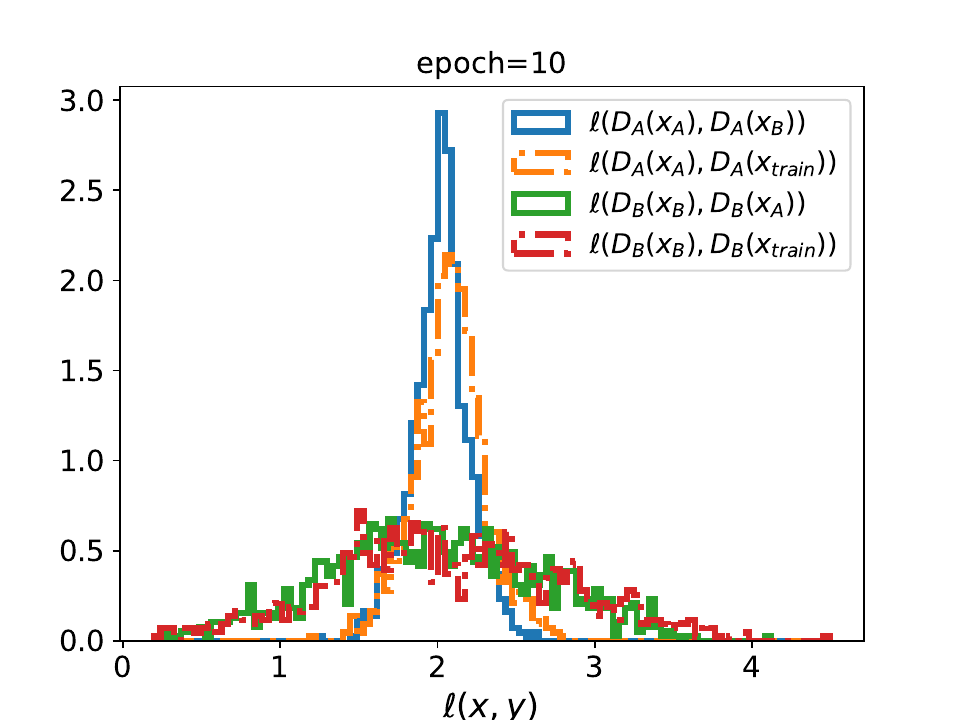}
    \includegraphics[width=0.8\linewidth]{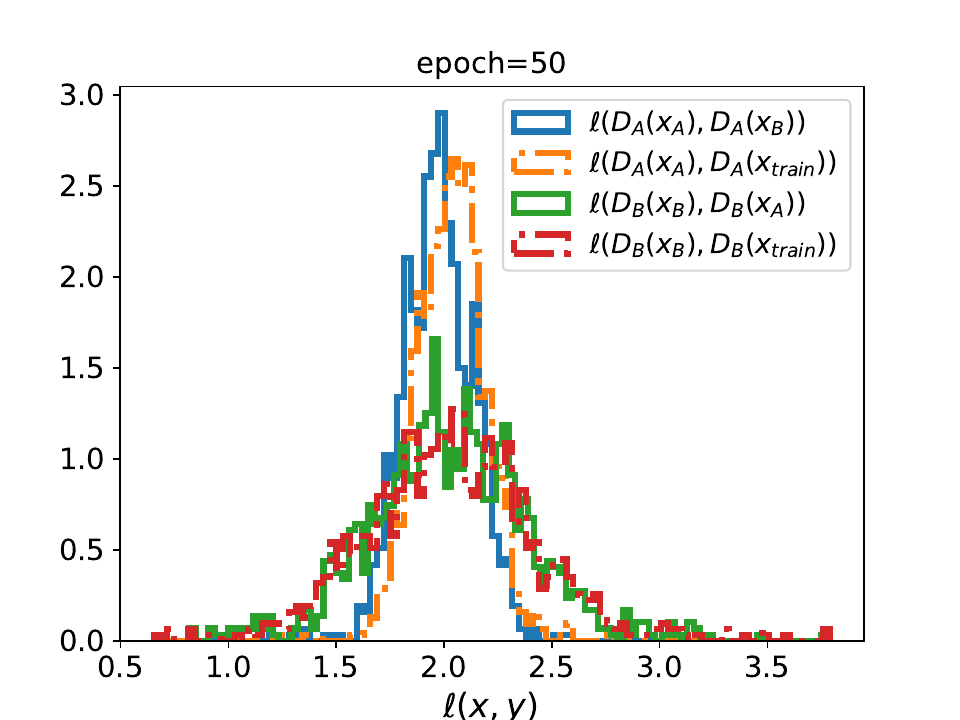}
    \includegraphics[width=0.8\linewidth]{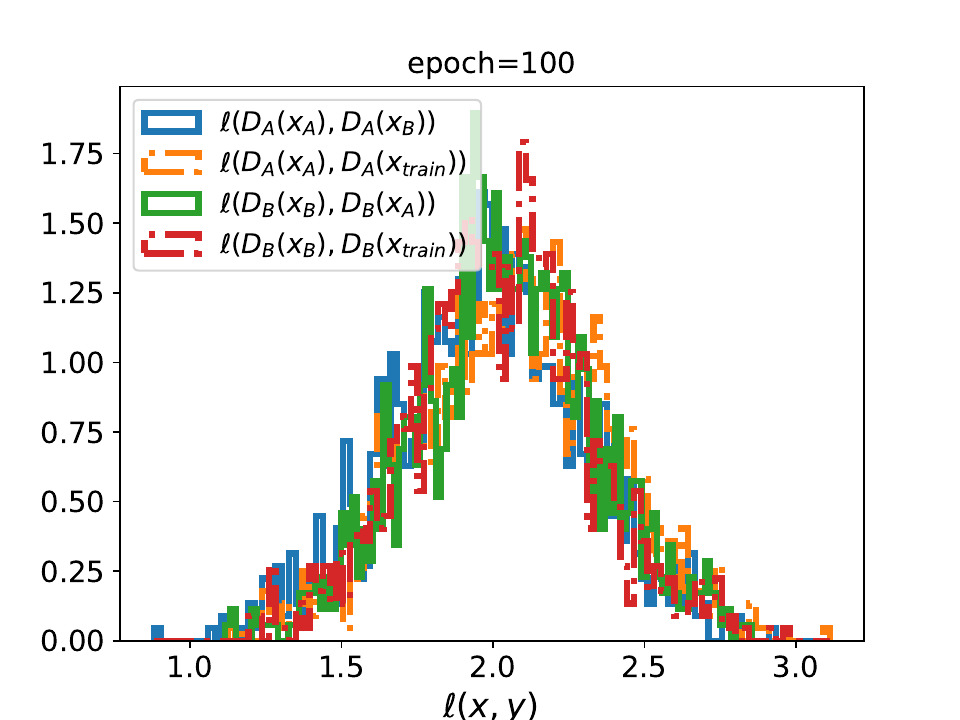}
\caption{Evolution of the hinge loss (Equation~\ref{eq-hinge-loss-ab}) for the "discriminator test" across various training epochs for models "A" and "B". Here, \(D_M(\bm{x}_{M^\prime})\) represents the output of the discriminator of model \(M\) applied to samples generated by the generator of model \(M^\prime\), while \(\bm{x}_{\text{train}}\) refers to samples from the respective training dataset used for both models. All models were trained using datasets consisting of $N=10^5$ images.}
\label{fig-gan-hinge-loss-evol}
\end{figure}

Figure~\ref{fig-gan-hinge-loss-evol} shows the results of the "discriminator test" at various optimization stages (\textit{epochs}) for models "A" and "B", both trained on datasets of size $N=10^5$, using $1,000$ samples of $\bm{z}$.
Some key observations include:
\begin{itemize}
    \item {Initial state (\textit{epoch} = 0)}: The discriminators \(D_A\) and \(D_B\), along with their associated generators, are clearly under-optimized, as the mean values of \(\ell\) are far from the theoretical target of 2. Additionally, for model "A", the distributions of \(\ell(D_A(\bm{x}_A), D_A(\bm{x}_B))\) and \(\ell(D_A(\bm{x}_A), D_A(\bm{x}_{\text{train}}))\) are identical, and a similar pattern is observed for model "B". However, the mean values not only differ from the target value, but they differ between the two models, likely due to the differences in their training datasets.
    \item {Progressive optimization (increasing \textit{epochs})}: As training progresses, all loss distributions gradually converge to a single distribution with a mean value of 2. This indicates that the discriminators and generators of both models are reaching their optimal states.
    When the four losses converge, it signifies that the discriminator for either model "A" or "B" no longer distinguishes between (i) samples from its associated generator, (ii) samples from the other model's generator, and (iii) training samples. 
\end{itemize}

Furthermore, we observe that the rate of convergence differs between the two models. If we focus solely on model "A", it might seem reasonable to stop optimization around \textit{epoch} 10, where \(\ell(D_A(\bm{x}_A), D_A(\bm{x}_B))\) and \(\ell(D_A(\bm{x}_A), D_A(\bm{x}_{\text{train}}))\) have converged satisfactorily to the target value of 2. However, this would prematurely halt optimization, as the stronger convergence criterion—the alignment of all four loss distributions—emerges only around \textit{epoch} 100. This highlights the utility of the two-models test in providing a more robust measure of convergence.

\begin{figure}
    \centering
	\includegraphics[width=0.7\linewidth]{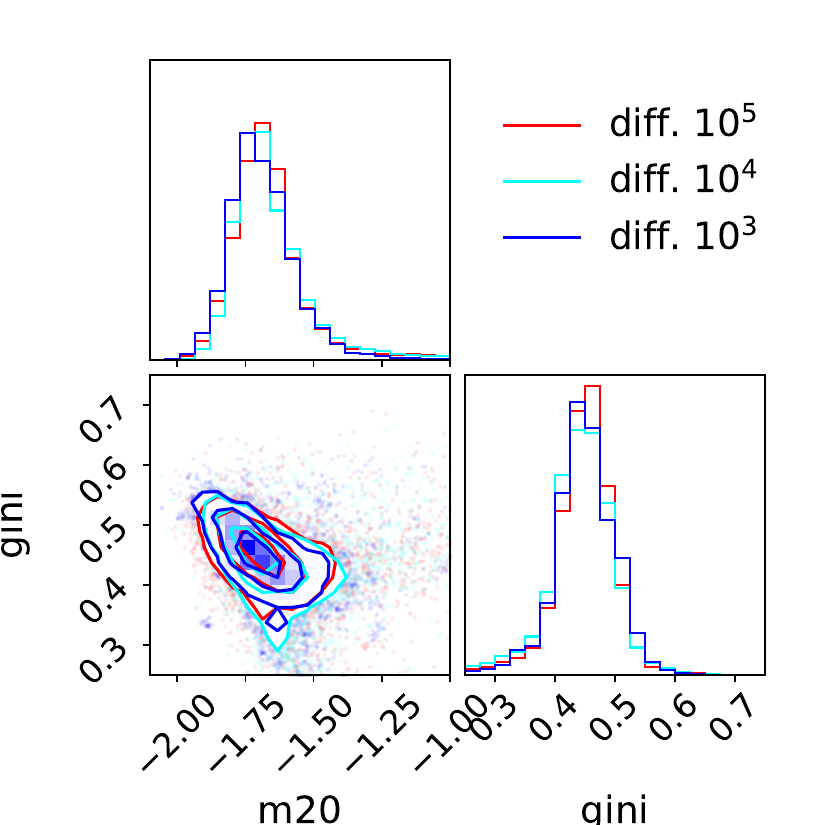}
	\caption{Some morphological coefficients (see Figure~\ref{fig-morpho-coeff}) computed using \texttt{light-weight-gan} models trained with different dataset sizes ($N=10^5, 10^4, 10^3$).}
	\label{fig-morpho-coeff-gan-comparison}
\end{figure}

Regarding the impact of reducing the training dataset size below $N=10^5$, we first analyze the morphological variables (Section~\ref{sec-Generated-Images}). As observed in Figure~\ref{fig-morpho-coeff-diff-comparison} for diffusion-based models, Figure~\ref{fig-morpho-coeff-gan-comparison} illustrates that the distributions remain comparable across training dataset sizes $N=\{10^5, 10^4, 10^3\}$. This observation would likely lead to the conclusion that it is possible to achieve good sampling of the underlying probability density with models trained on small batches of images.

\begin{figure}
    \centering
        \begin{subfigure}[b]{\columnwidth}
        \centering
        \includegraphics[width=0.8\linewidth]{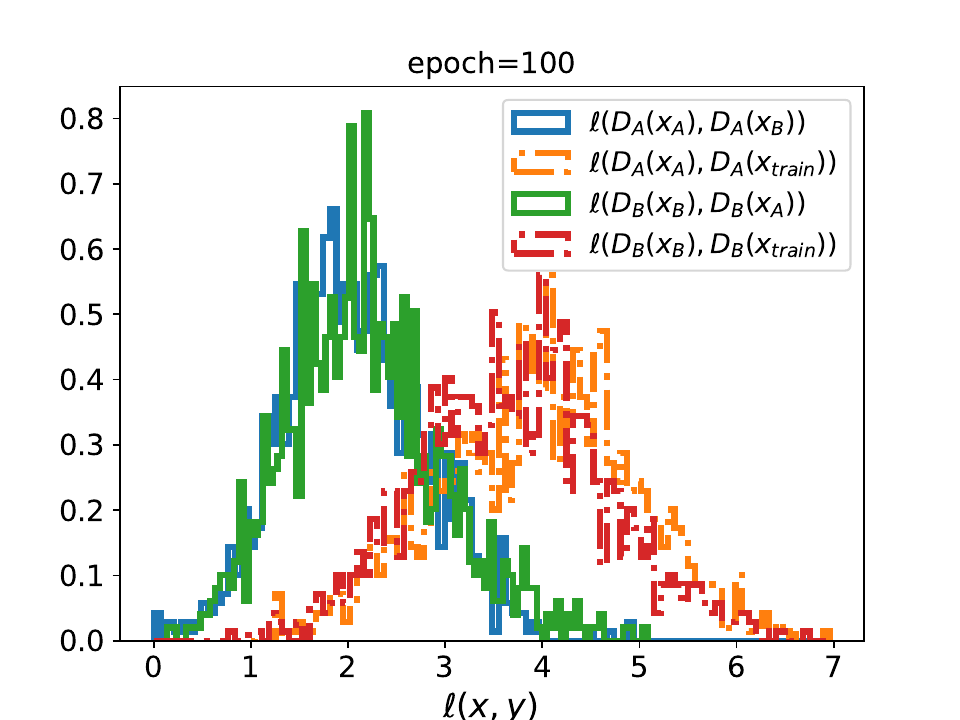}
        \caption{$N=10^4$}
        \end{subfigure}%
    \hfill
        \begin{subfigure}[b]{\columnwidth}
        \centering
        \includegraphics[width=0.8\linewidth]{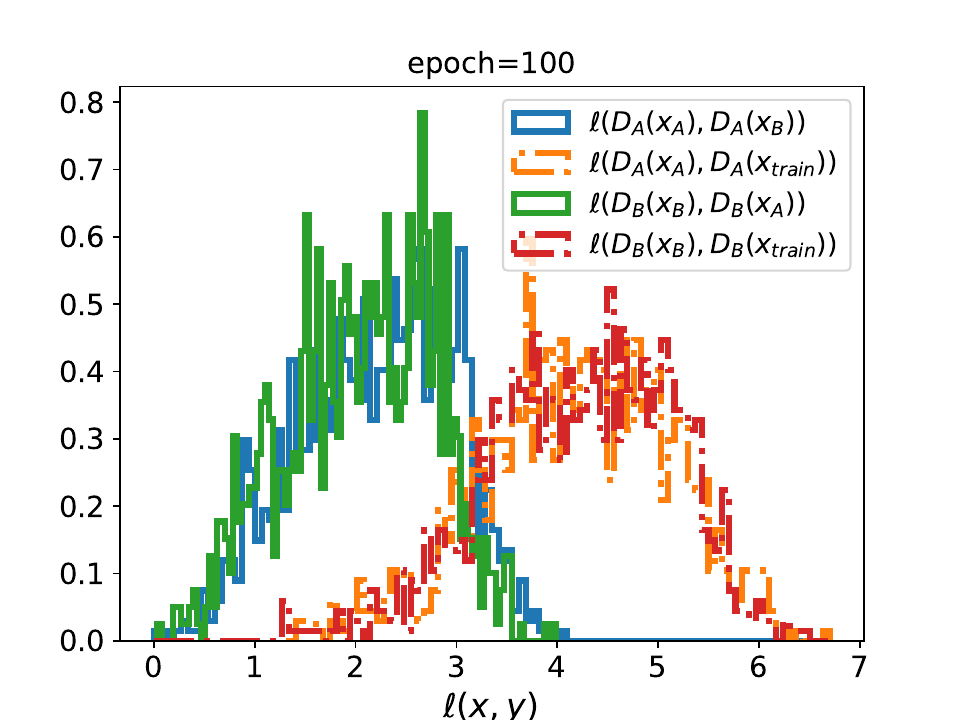}
        \caption{$N=10^3$}
        \end{subfigure}%
	\caption{Same legend as for Figure~\ref{fig-gan-hinge-loss-evol} but with different dataset sizes: $N=10^4$ (a) and $N=10^3$ (b).}
	\label{fig-gan-hinge-loss-epoch100-ntrain-10p3-10p4}
\end{figure}

However, Figure~\ref{fig-gan-hinge-loss-epoch100-ntrain-10p3-10p4} presents the results of the "discriminator test" for models trained with datasets of sizes $N=10^4$ (panel $(a)$) and $N=10^3$ (panel $(b)$), evaluated at \textit{epoch} 100. Readers are encouraged to compare these findings with the corresponding panel in Figure~\ref{fig-gan-hinge-loss-evol}. When training with $N=10^4$ or $N=10^3$, the results indicate that the discriminators are not entirely agnostic to the origin of the image samples, as evidenced by the distinct distributions of $\ell(D_A(\bm{x}_A), D_A(\bm{x}_B))$ and \(\ell(D_A(\bm{x}_A), D_A(\bm{x}_{\text{train}}))\) (and similarly for the $D_B$ discriminator). 
Moreover, the discrepancies in these loss distributions persist even when training continues beyond \textit{epoch} 100, suggesting that the observed divergences are not merely attributable to early convergence issues. It could be argued that these inconsistencies may arise from suboptimal hyperparameter choices in designing the \texttt{light-weight-gan} architecture used for the experiment, rather than solely from the limited dataset size. Notably, in the absence of convergence across the four loss distributions, there is no \textit{a priori} method to guide further investigations or remedial measures effectively.
\section{Conclusion}
\label{sec-conclusion}
In this study, we investigated the behavior of generative models applied to galaxy image generation, focusing on their performance under controlled experimental conditions. We { have presented three types of models as generic examples}: the \texttt{light-weight-gan} GAN-based model, the \texttt{Glow} flow-based model, and the diffusion model based on a \texttt{U-Net} denoiser (Section~\ref{sec-Exp-Models}). These models were trained and evaluated using identical datasets sourced from the SDSS survey (Section~\ref{sec-Dataset}).

Even with relatively small datasets, the generated images from all models appeared visually indistinguishable from the original ones (Figure~\ref{fig-Original-Glow-UNet-Gan-samples}). However, this study did not aim to achieve the highest possible image quality, as accomplished in prior works mentioned in the introduction. Instead, we focused on a different question: do models of the same type learn the same underlying probability distribution? The similarity in morphological features (Figure~\ref{fig-morpho-coeff})—used to assess whether the generated images are samples from the underlying data probability distribution function (pdf)—suggests that this is indeed the case. The main objective of this article was to approach this question from a novel perspective, revisiting the reliability of the generation processes. To this end, we designed experiments (Section~\ref{sec-experiment}) involving pairs of models trained on non-overlapping subsets of the data and seeded by identical latent variables, inspired by the methodology of \cite{kadkhodaie2024generalization}.

For the diffusion-based model (Section~\ref{sec-two_models-Diffusion}), our study of the cosine similarity metric between images revealed a transition from \textit{memorization} to \textit{generation} as the dataset size increased. This transition, rooted in the denoiser's capabilities, aligns with findings from \cite{kadkhodaie2024generalization}. Small datasets led to models that primarily memorized the training data, while larger datasets enabled generalization and the creation of novel images. This transition is essential for understanding the limits of such generative models and their ability to produce novel outputs. However, morphological distributions alone were insufficient to fully capture the nuances of this transition. It is worth noting that the forward and reverse processes in diffusion models are theoretically grounded in the Ornstein-Uhlenbeck equation and damped Langevin dynamics, adding robustness to their probabilistic framework.

For the flow-based \texttt{Glow} model (Section~\ref{sec-two_models-Flow}), while it generated samples with good morphological features, the cosine similarity metric indicated that the outputs were distinct from the training images, a positive outcome. However, two models trained on non-overlapping datasets of the same size produced different samples. Leveraging the bijective property of flow-based models, we proposed a consistency test called the "inversion test". Specifically, given two models "A" and "B," and a latent variables \(\bm{z} \sim \mathcal{N}(\bm{0}, \bm{1})\), we investigated whether latent variables \(\bm{z}^\prime = G_B^{-1}(G_A(\bm{z}))\) were identical to \(\bm{z}\) or at least shared the same pdf. Our findings showed that the latter was true only when the dataset size was sufficiently large, highlighting challenges faced by normalizing flows in capturing the true underlying data distribution with small datasets. This issue was also reflected in the degradation of morphological features with smaller dataset sizes.

For the \texttt{light-weight-gan} model (Section~\ref{sec-two_models-GAN}), the cosine similarity results were comparable to those of the flow-based model, but the GAN architecture lacks bijectivity. This distinction is inherent to the design of GANs. Nonetheless, we proposed a consistency test based on the hinge loss in the context of the "two-models" scenario. Using losses such as \(\ell(D_M(\bm{x}_M), D_M(\bm{x}_{M^\prime}))\) and \(\ell(D_M(\bm{x}_M), D_M(\bm{x}_{\text{train}}))\), where \(D_M\) is the discriminator of model \(M\) and $\bm{x}$ are either generated by a model or issued from the training datasets, we assessed whether the generators successfully learned the underlying data distribution. As training progressed, the "discriminator test" showed convergence of the mean loss to the theoretical value of 2, suggesting successful learning. However, discrepancies in the discriminator loss distributions, particularly with smaller datasets, revealed differences in generalization capabilities. Models trained on smaller datasets exhibited poorer generalization, even though their morphological features remained consistent, emphasizing the need for evaluations beyond morphological distributions.

In conclusion, while modern generative models can produce high-quality galaxy images with reasonably sized datasets, much remains to be explored regarding the mechanisms governing their learning processes and the factors influencing their generalization.  
{
As a key takeaway, we first highlight the necessity of sufficiently large training datasets to ensure effective learning. The exact amount of data required is not yet settled and remains an open research question, but it certainly depends on the image resolution, the number of model parameters, and the model architecture. Investigating different strategies to improve generalization without requiring massive datasets remains an open question. In future work, one could explore for instance restricting the diversity of galaxies in the dataset or leveraging pre-trained models on general-purpose images to enhance training efficiency.  
Secondly, we emphasize the importance of evaluations beyond morphological distributions to properly assess whether model optimization is effective. The "two-models" methodology proved to be a valuable approach, enabling us to assess both the similarity of generated images to real-world data and the consistency of the learned probability distributions. This framework has been a key element of our study.  
Finally, we argue that score-based diffusion models offer a direct way to test whether they operate in the generalization regime.  

We believe that the methods developed here are not limited to generating images of galaxies or to the specific models we have used.  
}

%
\section*{Acknowledgements}
We thank Prafulla Dhariwal who have accepted the reproduction of Figure 2 of  \cite{Kingma2018} (see Figure~\ref{fig-Glow-archi}). We thank Kim Seonghyeon for fruitful discussion on \texttt{Glow} model implementation. Similarly, we are grateful to Zahra Kadkhodaie and Florentin Guth for introducing me to their article. We acknowledge the use of  Nvidia’s V100 resources from French GENCI–IDRIS (Grant 2024-AD010413957R1).

\section*{Data Availability}
Codes, trained models and results, and the complete dataset of $250,000$ galaxy images are publicly available on request to the corresponding author. A GitHub repository is also accompanying this article, see details at \url{https://github.com/jecampagne/galaxy-gen-model-compagnon}. 

\bibliographystyle{aasjournal}
\bibliography{references}  

\bsp	
\label{lastpage}
\end{document}